\documentclass{aa}  
\usepackage{graphicx}
\usepackage{txfonts}
\usepackage{tabularx}
\usepackage{amsfonts}
\usepackage{bbold}
\usepackage{color}
\usepackage{transparent}
\usepackage{hyperref}
\usepackage{transparent}
\usepackage{rotating}
\usepackage{caption}
\usepackage{subcaption}

\usepackage{graphicx}	% Including figure files
\usepackage{amsmath}	% Advanced maths commands
\usepackage{amssymb}	% Extra maths symbols
\usepackage{tablefootnote}
\usepackage[flushleft]{threeparttable}
\usepackage{dblfloatfix} 
\usepackage{siunitx}
\usepackage{amsmath}

\usepackage{physics}

\usepackage{natbib}
\bibpunct{(}{)}{;}{a}{}{,} 

\usepackage[markup=nocolor]{changes}

\newcommand*{\eg}{e.g.\@\xspace}
\newcommand*{\ie}{i.e.\@\xspace}
\newcommand*{\aka}{a.k.a. \@\xspace}
\newcommand*\diff{\mathop{}\!\mathrm{d}}
\newcommand{\mystar}{{\fontfamily{lmr}\selectfont$\star$}}
\newcommand*{\msun}{$M_{\odot}$\@\xspace}

\newcommand*{\vac}{\texttt{MPI-AMRVAC}\@\xspace}
\begin{document} 

   \title{Wind morphology around cool evolved stars in binaries}

   \subtitle{The case of slowly accelerating oxygen-rich outflows}

   \author{I. El Mellah
          \inst{1,3}
         \and
          J. Bolte
          \inst{2} 
          \and
          L. Decin
          \inst{2}
          \and
          W. Homan
          \inst{2}
          \and
          R. Keppens
          \inst{1}
          }

   \institute{Centre for mathematical Plasma Astrophysics, 
   			 Department of Mathematics, KU Leuven, 
   			 Celestijnenlaan 200B, 3001 Leuven, Belgium
         \and
         	Institute of Astronomy, KU Leuven, Celestijnenlaan 200D B2401, 3001 Leuven, Belgium
         	\and
         	\email{ileyk.elmellah@kuleuven.be}
             }

   \date{Received ...; accepted ...}

% \abstract{}{}{}{}{} 
% 5 {} token are mandatory
  
  \abstract
   {The late evolutionary phase of low and intermediate-mass stars is strongly constrained by their mass-loss rate, which is orders of magnitude higher than during the main sequence. The  wind surrounding these cool expanded stars frequently shows non-spherical symmetry, thought to be due to an unseen companion orbiting the donor star. The imprints left in the outflow carry information on the companion but also on the launching mechanism of these dust-driven winds.}
  {We study the morphology of the circumbinary envelope and identify the conditions of formation of a wind-captured disk around the companion. Long-term orbital changes induced by mass-loss and mass transfer to the secondary are also investigated. We pay particular attention to oxygen-rich i.e. slowly accelerating outflows, in order to look for systematic differences between the dynamics of the wind around carbon and oxygen-rich asymptotic giant branch (AGB) stars.}
   {We present a model based on a reduced number of dimensionless parameters to connect the wind morphology to the properties of the underlying binary system. Thanks to the high performance code \vac, we run an extensive set of 70 three-dimensional hydrodynamics simulations of a progressively accelerating wind propagating in the Roche potential formed by a mass-loosing evolved star in orbit with a main sequence companion. The highly adaptive mesh refinement we use enables us to resolve the flow structure both in the immediate vicinity of the secondary, where bow shocks, outflows and wind-captured disks form, and up to 40 orbital separations, where spiral arms, arcs and equatorial density enhancements develop.}
   {When the companion is deeply engulfed in the wind, the lower terminal wind speeds and more progressive wind acceleration around oxygen-rich AGB stars make them more prone than carbon-rich AGB stars to display more disturbed outflows, a disk-like structure around the companion and a wind concentrated in the orbital plane. In these configurations, a large fraction of the wind is captured by the companion which leads to a significant shrinking of the orbit over the mass-loss timescale, if the donor star is at least a few times more massive than its companion. In the other cases, an increase of the orbital separation is to be expected, though at a rate lower than the mass-loss rate of the donor star. Provided the companion has a mass of at least a tenth of the mass of the donor star, it can compress the wind in the orbital plane up to large distances.}
   {The grid of models we compute covers a wide scope of configurations in function of the dust chemical content, the terminal wind speed relative to the orbital speed, the extension of the dust condensation region around the cool evolved star and the mass ratio. It provides a convenient frame of reference to interpret high-resolution maps of the outflows surrounding cool evolved stars.}

   \keywords{stars: AGB and post-AGB binaries -- (stars:) binaries: general -- stars: winds, outflows -- stars: evolution -- accretion, accretion discs -- methods: numerical}

   \maketitle
%
%________________________________________________________________

\section{Introduction}

% A. COOL EVOLVED STARS

% A.1. DEFINITION AND CONNECTION W/ BIG PICTURE

As they evolve beyond the main sequence, low and intermediate-mass stars of initial masses ranging from 0.8 to 8\msun are bound to expand and cool down. Along the red giant branch (RGB), the effective gravity in the outermost layers drops and the mass-loss rate increases \citep{Groenewegen2012}. During the subsequent asymptotic giant branch (AGB) phase, the mass-loss rate peaks at $10^{-7}$ to $10^{-5}$\msun yr$^{-1}$, with terminal wind speeds of 5 to 20\,km s$^{-1}$ \citep{Knapp1998,Habing2003,Herwig2005,Ramstedt2009}. By the time they eventually become white dwarfs, AGB stars lost 35 to 85\% of their mass \citep{Marshall2004}. The stars with an initial mass above $\sim$8\msun undergo  a phase where they manifest as red supergiants \citep[RSG,][]{Levesque2010}.

% A.2. WIND LAUNCHING

Through their winds, cool evolved stars represent a major source of dust and molecular enrichment for the interstellar medium \citep{Groenewegen2002}. Although the wind launching mechanism remains largely unknown for RSGs, a dust-driven model proved successful for AGB stars \citep[see][for a recent and comprehensive review]{Hofner2018}. The $\kappa$-mechanism excites radial pulsations which shoot out material and provoke the formation of internal shocks within a couple of stellar radii. The compressed gas cools down downstream the shocks until it reaches temperatures of 1\,500 to 1\,800K, low enough to trigger the condensation of gaseous species into dust grains which absorb the stellar radiation, accelerate and drag the ambient gas \citep{Liljegren2016,Freytag2017}. During their post-main sequence evolution, low and intermediate-mass stars experience several dredge-up events which bring freshly formed carbon to the outer envelope and can turn originally oxygen-rich (O-rich) stars into carbon-rich (C-rich) stars \citep{Straniero1997a,Herwig2004} with dramatic consequences on the dust chemical content. Since the latter sets the dust opacity, the wind acceleration profile strongly differs around O-rich and C-rich AGB stars: the low opacity of dust grains formed from O-rich gaseous environments leads to a much more progressive wind acceleration around O-rich AGB stars, where the wind reaches its terminal speed several tens of stellar radii away from the donor star \citep{Decin2010a,Decin2018}. A full understanding of dust-driven winds emitted by cool evolved stars requires a radiative hydro-chemical model of dust grains growth in an environment generally out of thermodynamic equilibrium \citep{Boulangier2019}. %, a challenge whose increasing appeal owns much to game-changing observational devices.
   
% B. WIND MORPHOLOGY

% B.1. INSTRUMENTS: WHAT ARE WE LOOKING AT?

High spatial and spectral resolution instruments have shed an unprecedented light on the complexity of the astrochemistry at work in these cool winds. While the Hubble space telescope ushered in a new era of high-resolution imaging in the optical wavelength range, the infrared space telescope Spitzer revealed how diverse the wind properties can be around cool evolved stars in the Galaxy but also in the nearby small and large Magellanic clouds \citep{DellAgli2015,Yang2018}. The Hubble space telescope captures the light scattered by the dust present in large quantities in the inner regions of the wind but molecular line emission also contains invaluable information. The powerful spectrometers aboard Herschel  and the (sub)millimeter interferometer ALMA granted us access to a plethora of rotational and vibrational lines originating up to several thousands of astronomical units (au) away from the cool evolved star. Thanks to multi-channel molecular emission maps, the kinematic structure of the circumstellar flow can be characterized, opening the door to a full 3D reconstruction of the wind \citep{Decin2018,Decin2020}.

% Herschel \citep{Pilbratt2010}

% B.2. OBSERVATIONS

Significant non-spherical features in the wind of cool evolved stars have been identified thanks to these instruments, but also in the shape of planetary nebulae, which are thought to be the descendants of RGB and AGB stars \citep{Shklovsky1956}. Recurrent axisymmetric patterns appear around cool evolved stars such as compressed structures \citep[disks or tori,][]{Bujarrabal2016,Kamath2016}, polar cavities and spirals \citep{Homan2018}, but also arcs \citep{Decin2020}. Although alternative explanations have been proposed \citep{Nordhaus2006,Chita2008}, the community has been progressively more inclined to attribute these features to the presence of an unseen companion. The reason of this shift in opinion stems from observations as well as new population synthesis results \citep{Moe2017}. In a few systems where non-spherical imprints are observed in the circumstellar environments, robust hints in favor of the presence of a companion have emerged. In these systems, it is still questioned though whether the companion is close enough to be responsible for producing these non-spherical features or if there is a third undetected object on a close orbit with the donor star. Around RGB stars with periodically modulated luminosities, simultaneous monitoring of the Doppler shift and of the light curve proved that these variations were due to the ellipsoidal deformation of the star due to the presence of a companion on a close orbit \citep[sequence E long period variables,][]{Nicholls2010}. Furthermore, the complex morphology of the AGB wind in the Mira AB system has been shown to be partly due to a companion on a $\sim$500 years orbit \citep{Ramstedt2014}.

%Furthermore, half of solar-mass stars in the solar neighbourhood have been found to be in a multiple system \citep{Raghavan2010}, in agreement with population synthesis studies \citep{Izzard2018}. TRUE? Binary population synthesis models suggest that a large fraction of AGB stars are in binary or multiple systems (SANA+12 REF BUT FOR LOW-MASS STARS?).

% B.3. C /= O

% Justtanont2015: Oxygen isotopic ratio as a tracer of the initial mass of an AGB star.

% C. BINARIES

The impact of binarity on stellar evolution can not be overstated. Interactions in a close binary system modify the chemical stratification within the star and alter the evolution of its spin via mass and angular momentum exchanges with the orbiting companion \citep{DeMarco2017a}. It has been shown to be a key-ingredient to understand Barium stars \citep{Bidelman1951}, SNe Ia as descendants of symbiotic binaries \citep{Claeys2014}, Carbon and s-process enhanced metal-poor stars \citep{Abate2013} and  blue stragglers in old open clusters \citep{Jofre2016}. Classic Roche lobe overflow (RLOF) and wind accretion have been shown to be the two limit mass transfer mechanisms of generally more complex configurations. Numerical hydrodynamics simulations enabled hybrid regimes such as wind-RLOF to be identified \citep{Mohamed2007}, which turned out to be decisive to reconcile models and observations. They revealed enhanced mass transfer rates between the donor star, also called the primary, and the accretor, also called the secondary \citep{Abate2013}, and the formation of wind-captured disks around the secondary \citep{HuarteEspinosa:2012wq,DeVal-Borro2017}. 

Grid-based and smooth particle hydrodynamics numerical simulations also reproduced large scale features left by the secondary in the wind of the donor star, such as arcs, spirals and circumbinary disks, and evaluated the impact of the inclination of the orbit with the line-of-sight \citep{Mastrodemos1999,Liu2017,Chen2018,Saladino2019,Kim2019}. However, the slow acceleration of O-rich outflows proved to be a challenging ingredient to account for, along with controlling the amount of kinetic energy per unit mass provided to the flow by the radiative pressure on dust grains. Resolving both the wind launching scale and the whole circumbinary outflow requires highly adaptive grids whose geometry matches the one of the outflow. As noted by \cite{Chen2017}, spherical grids represent good candidates to move forwards. This is why in this work, we developed a numerical framework based on a radially stretched spherical mesh centered on the donor star. It highly facilitates the treatment of the wind launching, guarantees the conservation of angular momentum and limits numerical artifacts introduced by the discretization of the equations (\eg numerical diffusivity). It also enables us to capture both the initial deviation of the wind by the secondary and the shaping of the circumbinary envelope up to several 10 orbital separations at a computational cost so low that we can cover a wide range of realistic sets of parameters. With the dimensionless parametrization we introduce, we can parametrically explore how the properties of the dust-driven acceleration and of the accreting companion determines the morphology of the outflow. 

% Since the wind launching efficiency heavily depends on dust and its chemical content, the morphology of the circumbinary envelope in binary systems containing a cool evolved star provides new insights on the late evolutionary stages of low and intermediate-mass stars and on dust growth around cool evolved stars.

In Section\,\ref{sec:model}, we present a model based on an empirical wind acceleration so as to reproduce a predefined velocity profile. We also identify the dimensionless parameters which determine the shape of the hydrodynamics solutions we solve in our numerical setup. In Section\,\ref{sec:results}, we report on the main results of our comprehensive exploration of 70 configurations representative of C-rich and O-rich AGB stars with an orbiting companion. It provides a grid of models spanning all the possible configurations whose main properties are summarized and discussed in Section\,\ref{sec:discussion}.

% Limitations of current proxies: some systems show jets (Claudia paladini?) => disk-fed accretion onto the secondary, which cannot be described through the wind accretion formula provided by BHL. 

% Wind accretion studied numerically: state-of-the art. Binaries accreting from ISM \citep{Antoni2019}, formation of wind-captured disks during the common envelope phase \citep{Murguia-Berthier2017}.

% But many observed jets in systems too wide for the donor star to fill its Roche lobe (REF?) —> wind accretion. For long, we used BHL to compute mass accretion rates due to wind accretion. Wind-captured disks: wind-RLOF produces enhanced mass transfer rates (Saladino2018) since lower fraction of mass-loss captured but much larger mass-loss rate available so it can compensate, also in high mass X-ray binaries (ElMellah2018a) => misleading BHL when wind speed comparable to orbital speed, which is the case in AGB binaries (and in HMXB), although still used in binary population-synthesis calculations. 

% ------------------------------------------------
\section{Model}
\label{sec:model}
% ------------------------------------------------

% - - - - - - - - - - - - - - - - - - - - - - - - 
\subsection{Parametrized $\beta$-wind}
\label{sec:para_wind}
% - - - - - - - - - - - - - - - - - - - - - - - - 

In this section, we describe how the wind is accelerated from the donor star. The wind launching does not depend on time and stellar pulsations are not included in the present model \citep[see][for a model with periodic modulations of the mass-loss rate through the radial outflow velocity]{Chen2017}. We use the classic $\beta$-wind velocity profile as a parametrization to compute the corresponding radiative acceleration term for an isolated star. This method guarantees that we control both the terminal speed and how quickly the wind reaches it, contrary to previously used approaches which parametrize directly the acceleration term: alternatives include winds undergoing reverted gravity \citep{Kim2019} or the free-wind model where radiation counterbalances exactly gravity, leading to a constant wind speed \citep{Liu2017} or a thermally-driven wind when thermal pressure is included \citep{Saladino2019}. An enforced $\beta$ velocity profile alleviates the full radiative-hydrochemical computation required to determine the acceleration profile, allowing us to explore the impact of a wide range of realistic acceleration profiles at an affordable computational cost. % For a more physical treatment of the wind launching around AGB stars, the reader is invited to look up \cite{Chen2017}.  

\subsubsection{Presureless test case}

We first validate our approach in the simplemost context: a pressureless purely radial wind in spherical geometry undergoing gravity and an outwards radiative force whose large amplitude justifies that we neglect the pressure force which drives thermal winds \citep{Lamers1999}. These radiatively-driven winds are well described by a velocity profile called the $\beta$-law \citep{Puls2008}:
\begin{equation}
    \label{eq:beta-law}
    v_{\beta}(r)=v_{\infty}(1-R_0/r)^{\beta}
\end{equation}
where $R_0$ is the distance to the stellar center where acceleration triggers, $v_{\infty}$ is the terminal wind speed and $\beta$ is a positive exponent whose value determines how quickly the wind reaches its terminal speed, with a more progressive acceleration for a larger value of $\beta$. Around hot stars, \cite{Lucy1970} and \cite{Castor1975} designed a model which describes how the resonant line absorption of the many UV photons available by partly ionized metal ions provides momentum to the outer envelopes of the stellar atmosphere. These winds are said to be line-driven and a physically-motivated analysis leads to the velocity profile in equation\,\eqref{eq:beta-law}. Around AGB stars, wind launching is believed to be made possible thanks to pulsations which lift up material up to a distance, the dust condensation radius $R_d$, where the temperature is low enough and the density high enough owing to the shocks for dust to form. The subsequent radiative pressure force of the continuum on the dust grains accelerate them and once they redistribute their momentum to the ambient gas through drag, a wind is launched \citep{Hofner2018}. Around other types of cool evolved stars such as RSGs, the wind launching mechanism remains largely unknown, although the observed velocity profiles still reasonably match the aforementioned $\beta$-law \citep{Decin2006}.

The terminal wind speed of radiatively-driven winds increases with the escape speed, which leads to much larger terminal speeds for hot stars ($\sim$ 1\,000\,km s$^{-1}$) than cool evolved stars ($\sim$ 10\,km s$^{-1}$). The acceleration starts very close from the photosphere around hot stars while it only starts at the dust condensation radius for winds of AGB stars. Depending on the stellar metallicity, the $\beta$ exponent ranges between 0.5 and 2.5 for hot stars \citep{Sander2017} while it varies much more for AGB stars depending on the chemical content. C-rich AGB stars lead to dust grains of high opacity, strongly accelerated by the stellar radiative field. The terminal speed is reached within a few stellar radii, which yields a $\beta$ exponent as low as 0.1 \citep{Decin2015a}. Due to the low opacity of the dust grains formed in their midst, winds surrounding O-rich AGB stars accelerate in a much more progressive way, with $\beta$ exponents as high as 5 \citep{Khouri2014}.

The implementation of this configuration in a uni-dimensional numerical setup leads to the stable profile represented in Figure\,\ref{fig:vel_prof_1D} (blue solid line). The initial condition is a flat density profile with zero-speed and the analytic solution is set at the inner boundary condition located at 1.01$R_0$. The acceleration source term is given by $ v_{\beta}\diff v_{\beta}$. This preliminary benchmark confirms the validity of the solving schemes used later on in the full 3D setup described in \ref{sec:num_mod}.

% The typical acceleration profiles and the physical meaning of the main parameters in the $\beta$-law are discussed in more detail in Appendix\,\ref{app:app1}.

\subsubsection{Modified $\beta$-wind}
\label{sec:mod_bet}

\begin{figure}[!b]
\centering
\includegraphics[width=0.99\columnwidth]{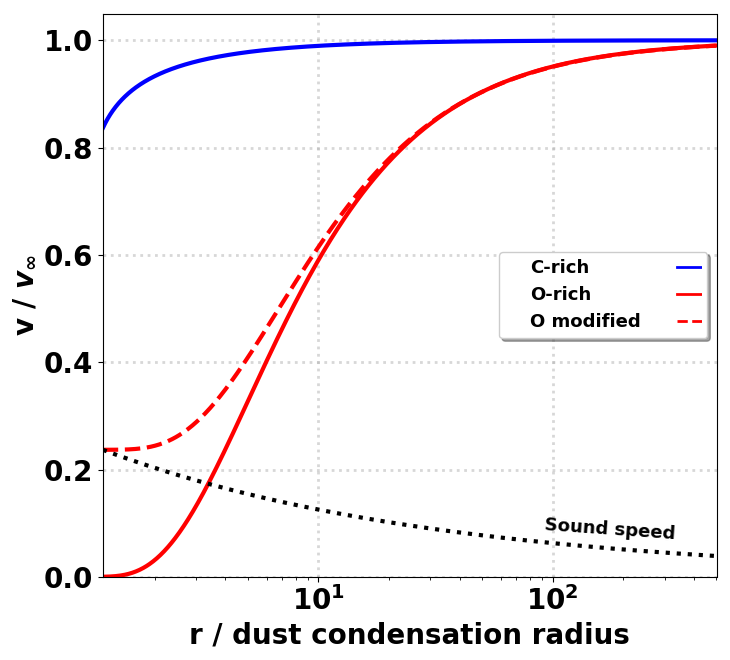}
\caption{Velocity profiles for a C-rich AGB star (solid blue line, $\beta=0.1$) and for an O-rich AGB star (solid red line, $\beta=5$), from 1.2 to 500 dust condensation radii. The red dashed line represents the modified $\beta$-law such as the wind is launched at a velocity slightly above the sound speed at the inner edge of the simulation space ($\beta'\sim6.7$). The sound speed profile is the dotted black line and we assumed a fiducial terminal wind speed 4 times higher than the sound speed at the dust condensation radius.}
\label{fig:vel_prof_1D}
\end{figure} 

A specific difficulty for numerical simulations of winds around O-rich AGB stars arises from the low wind speed up to large distances. For C-rich AGB stars, the sonic point at radius $R_s$ is very close from the dust condensation radius which ensures that the wind can safely be injected from $r=R_d$ at the sound speed or slightly above, as it has been done in previous simulations \citep{Kim2019}. It prevents spurious reflections of acoustic waves at the injection border of the simulation box. For O-rich AGB stars though, such a procedure requires more caution since the wind might remain subsonic up to several stellar radii. 

In case the $\beta$-law in equation\,\eqref{eq:beta-law} gives a subsonic wind speed at the inner edge of the simulation space, set at 1.2$R_d$, we need to modify the $\beta$ velocity profile such as the wind starts a few 0.01\% above the local sound speed $c_{s,s}$:
\begin{equation}
    \label{eq:beta-law_mod}
    v_{\beta,mod}(r)=c_{s,s}+(v_{\infty}-c_{s,s})(1-R_d/r)^{\beta'}
\end{equation}
where $\beta'$ is computed such that the terminal radius remains unchanged. This fix is generally needed for O-rich stars but not for C-rich stars. We compute the local sound speed assuming a radial temperature profile $T\propto r^{-0.6}$, in agreement with the observations, a temperature at the dust condensation radius of 1,500K and a fiducial mean molecular weight of 1 proton mass. An additional argument in favor of the legitimacy of this approach comes from simulations of the flow structure in the innermost parts, between the stellar photosphere and the dust condensation radius. For instance, \cite{Freytag2017} performed simulations of the outer envelopes of AGB stars and found characteristic turbulent speeds associated with Mach numbers of 2 to 3. A modified $\beta$ velocity profile is represented in Figure\,\ref{fig:vel_prof_1D} (red dashed line) where it can be seen that even for very slowly accelerating winds, the discrepancy with the classic $\beta$-law (red solid line) remains moderate. Hereafter, we write $v_{\beta}$ both for the standard and modified $\beta$-law, depending on whether a correction is needed or not.

% - - - - - - - - - - - - - - - - - - - - - - - - 
\subsection{Binary system and equations}
\label{sec:bin_eq}
% - - - - - - - - - - - - - - - - - - - - - - - - 

The aforementioned launching procedure guarantees that the right amount of kinetic energy per unit mass is progressively injected into the wind, given a certain predefined velocity profile. The modeling of the full binary can now be described. We work in the frame co-rotating with the two bodies, at the orbital angular speed $\Omega$, neglecting any eccentricity of the orbit. To limit the number of parameters, the star is assumed to have no spin since its significant radial expansion leads to a spin angular speed much lower than the orbital angular speed $\Omega$. Moreover, \cite{Saladino2019} studied the secular evolution of the spin of the AGB donors in binaries due to mass-loss and tidal interactions and concluded that the stellar spin was negligible in the AGB phase for orbital separations larger than 5 au. The strong dependency of the synchronization timescale on the ratio of the stellar radius to the orbital separation might, however, lead to fast spinning-up of the donor star when it gets close to fill its Roche lobe \citep{Zahn1977}. We also discard gas self-gravity and the feedback of the wind on the stellar masses and orbits, negligible over the duration of our simulations.

The fundamental equations of hydrodynamics which determine the evolution of the stellar wind are the continuity equation and the conservation of linear momentum:
\begin{equation}
\label{eq:eq1}
\begin{aligned}
\partial _t \rho + \boldsymbol{\nabla} \cdot \left( \rho \mathbf{v} \right) = 0
\end{aligned}
\end{equation}
\begin{equation}
\label{eq:eq2}
\begin{aligned}
\partial _t \left( \rho \mathbf{v} \right) + \boldsymbol {\nabla} \cdot \left( \rho \mathbf{v} \otimes \mathbf{v} + P \mathbb{1} \right) = &- \rho \frac{GM_2}{r_2^3}\boldsymbol{r_2} \\ 
&- \rho \boldsymbol{\Omega}\wedge\left(\boldsymbol{\Omega}\wedge\boldsymbol{r}\right) - 2\boldsymbol{\Omega}\wedge\rho\mathbf{v} \\
&+ \rho v_{\beta}\diff_r v_{\beta} \frac{\boldsymbol{r_1}}{r_1}
\end{aligned}
\end{equation}
where $\rho$, $\mathbf{v}$ and $P$ are the mass density, the velocity vector and the pressure respectively, while $G$ is the gravitational constant. $\mathbb{1}$ is the diagonal unity 3 by 3 matrix, $\otimes$ is the dyadic product and $\wedge$ is the cross product. The subscripts $1$ and $2$ refer to the donor star and to the secondary object respectively, with $M_{i}$ the mass of body $i$ and $r_i$ the position vector with respect to body $i$. Finally, $\boldsymbol{r}$ is the position vector with respect to the center of mass of the 2 bodies. The first term on the right hand side of equation\,\eqref{eq:eq2} represents the gravitational influence of the secondary object and the second and third terms are the centrifugal and Coriolis forces. The last term includes the gravitational attraction of the donor star and the radiative acceleration, as described in Section\,\ref{sec:para_wind}, with $v_{\beta}$ given by equation\,\eqref{eq:beta-law} or \eqref{eq:beta-law_mod}. The orbital angular speed $\Omega$ is related to $M_1$, $M_2$ and the orbital separation $a$ via Kepler's 3$^{\text{rd}}$ law. The coupling of the gas with the dust is encapsulated in this radiative acceleration term. 

We need an additional equation to determine the evolution of the pressure entering equation\,\eqref{eq:eq2}. Ideally, this equation would account for the full thermodynamics of the problem and in particular, for the heating by the stellar radiative field and re-emiting dust grains \citep{Boulangier2018,Boulangier2019} and for the electron and molecular cooling mechanisms. Instead, we will rely on the following polytropic prescription to close the system of equations:
\begin{equation}
P=S\rho^{\gamma}
\label{eq:eq3}
\end{equation}
where $\gamma$ is the polytropic index and $S$ is a constant which can be related to the sound speed at the sonic point $c_{s,s}$ and the density at the sonic point $\rho_s$ through $S=c_{s,s}^2/\left(\gamma \rho_s^{\gamma-1}\right)$. 

Physically, a polytropic index of $5/3$ would mean that we neglect any net heating/cooling (adiabatic hypothesis) while a polytropic index of $1$ would mean that heating and cooling are efficient enough to counterbalance any change of internal energy due to expansion or compression respectively (isothermal hypothesis). To determine the suitable polytropic index to reproduce the observed temperature profile of $T\propto r^{-0.6}$, we follow this reasoning: due to the polytropic assumption and the ideal gas law, $T\propto \rho^{\gamma-1}$. In addition, due to mass conservation in spherical geometry, when the wind speed has reached the terminal speed, $\rho\propto r^{-2}$. Consequently, a polytropic index $\gamma=1.3$ enables us to represent a realistic cooling and to match the observed temperature profile far from the star.

The wind is launched from a sphere of radius $R_{\text{in}}=1.2R_d$ with a density set by the sonic radius $R_s$ with respect to $R_{\text{in}}$ and a radial velocity given by the $\beta$-law, modified if needed. The toroidal component of the velocity vector is set by the absence of spin of the donor star in the inertial frame of the observer.

We rely on the following normalization, which reduces the number of parameters, needed to numerically compute the shape of the solutions, to 5:
\begin{itemize}
    \item \textit{length:} the condensation radius $R_d$,
    \item \textit{speed:} the orbital speed $a\Omega$,
    \item \textit{density:} the sonic point density $\rho_s$, linked to the mass-loss rate.
\end{itemize}

The 5 fundamental dimensionless shape parameters which appear in the equations after normalization and entirely determine all possible numerical solutions are:
\begin{itemize}
    \item the mass ratio $q=M_1/M_2$,
    \item the dust condensation radius filling factor $f=R_d/R_{R,1}$, with $R_{R,1}$ the Roche lobe radius of the primary given by \cite{Eggleton1983},
    \item the ratio of the terminal to the orbital speed $\eta=v_{\infty}/\left(a\Omega\right)$,
    \item the $\beta$ exponent which sets the steepness of the velocity profile,
    \item the ratio of the terminal speed to the sound speed at the dust condensation radius $v_{\infty}/c_{s,d}$.
\end{itemize}

To identify the main dependencies of the wind morphology on these parameters, we consider different values for these parameters, representative of AGB binaries. We take mass ratios of 1 and 10, for respectively an AGB star in orbit with a main sequence solar-type companion and with a low-mass star or a brown dwarf companion. The filling factor varies much from one system to another since both RGB and AGB donor stars can be large enough and/or the orbital separation can be small enough such that the dust condensation region (or even the star itself) might fill the Roche lobe. We work with a filling factor of 5\%, 20\% and 80\% to study the impact of the stellar expansion and/or the secular evolution of the orbital parameters on the wind dynamics. To distinguish between the quickly accelerated winds of C-rich AGB stars and the slowly accelerating winds of O-rich AGB stars, we consider $\beta$ exponents of 0.1 and 5 respectively. The terminal wind speed compared to the orbital speed takes 6 different values, 0.5, 0.8, 1.2, 2, 4 and 8. Given the limited influence of the flow temperature quantified through the last parameter, we set it to a realistic value and use $c_{s,d}=v_{\infty}/4$. This parametrization enables us to span all possible configurations with only 72 simulations whose numerical setup we now describe.

Typical values for the normalization units are the following. A stellar radius $R_{\text{\mystar}}\sim 1$ au is realistic for an AGB star \citep{Gobrecht2016}. Dust has been detected as close as $\sim$1 stellar radius above the stellar photosphere \citep[see the O-rich AGB star R Doradus,][]{Khouri2016}, although they might be dust seeds which are still unable to drive the wind and which only coated into high enough opacity grains further in the wind \citep{Hofner2016}. \cite{Gobrecht2016} reports a dust condensation radius of $\sim 3-4$ stellar radii around the O-rich AGB star IK Tau, but estimates as large as 15\,$R_{\text{\mystar}}$ exist \citep{Sargent2011}. The orbital speed $a\Omega$ varies widely, from 3\,km s$^{-1}$ for wide binaries containing an AGB donor with orbital periods of the order of a few 1\,000 years and a binary mass of the order of 1\msun, up to 15\,km s$^{-1}$ for a binary mass of 5\msun and an orbital period as low as 50 years \citep[for CW Leo,][]{Decin2015}. For a particular class of RGB stars which manifest themselves as sequence E long period variables, a companion is believed to be on a close orbit with typical orbital periods of 100 days and orbital speeds $a\Omega$ of $\sim$50\,km s$^{-1}$ \citep{Nicholls2010}. Finally, the stellar mass-loss rate $\dot{M}_1<0$ is used to set the density at the sonic point:
\begin{equation}
    \rho_s=\frac{\abs{\dot{M}_1}}{4\pi c_{s,s}R_s^2}=\frac{\abs{\dot{M}_1}}{4\pi c_{s,d}R_d^2}\left(\frac{R_d}{R_s}\right)^{1.7}
\end{equation}
where the last factor accounts for the radial temperature profile and ranges between 0.75 and 1 since $R_s$ ranges from $R_d$ (for a quickly accelerating wind) to 1.2$R_d$ (for a slowly accelerating wind corrected with the modified $\beta$-wind profile described in Section\,\ref{sec:mod_bet}). The exponent 1.7 comes from the assumed radial temperature profiles. For a realistic sound speed at the dust condensation radius of $\sim$3\,km s$^{-1}$, a mass-loss rate from 10$^{-7}$\,\msun yr$^{-1}$ to 10$^{-5}$\,\msun yr$^{-1}$ and a dust condensation radius between 2 and 5 times the typical stellar radius $R_{\text{\mystar}}\sim 1$ au, we obtain densities at the sonic point ranging between 2$\cdot$10$^{-16}$g cm$^{-3}$ and 2$\cdot$10$^{-13}$g cm$^{-3}$. As a reference, we report in Table\,\ref{tab:para} the physical parameters deduced from observations for the classic C-rich AGB star CW Leo (\aka IRC$+$10216) and its companion.
\begin{center}
\begin{table}[!t]
\caption{Parameters of the C-rich star CW Leo and its companion. In the top part of the table, the results from \cite{Decin2015} and \cite{Cernicharo2015} are reported, where both teams essentially disagreed on the orbital period P, leading to different orbital speeds. In the second part, we give the corresponding normalization units in cgs. In the last part, we compute the corresponding dimensionless parameters.}
\label{tab:para}
\centering
\begingroup
\setlength{\tabcolsep}{10pt} % Default value: 6pt
\renewcommand{\arraystretch}{1.3} % Default value: 1
\begin{tabularx}{0.82\columnwidth}{c|cc}
   & Decin$+$15 & Cernicharo$+$15 \\
  \hline
  M$_1$ & 4M$_{\odot}$  & 1.5M$_{\odot}$  \\
  M$_2$ & 0.6M$_{\odot}$  & 1.5M$_{\odot}$  \\ 
  R$_1$ & \multicolumn{2}{c}{3 au} \\
  R$_d$ & \multicolumn{2}{c}{5R$_1$} \\
  $\dot{M}_1$ & -1.5$\cdot$10$^{-5}$\msun yr$^{-1}$  & -2$\cdot$10$^{-5}$\msun yr$^{-1}$ \\
  v$_{\infty}$ & \multicolumn{2}{c}{14.5km s$^{-1}$} \\ 
  P & 55 years & 800 years \\
  a & 25 au & 100 au \\
  \hline
  R$_d$ & \multicolumn{2}{c}{2.2$\cdot$10$^{14}$cm} \\
  a$\Omega$ & 13$\cdot$10$^{5}$cm s$^{-1}$ & 4 $\cdot$10$^{5}$ cm s$^{-1}$ \\
  $\rho_s$ & 5$\cdot$10$^{-15}$ g cm$^{-3}$ & 7$\cdot$10$^{-15}$ g cm$^{-3}$\\
  \hline
  q & 7 & 1\\
  $f$ & \multicolumn{2}{c}{$\sim$ 20\%} \\
  $\beta$ & \multicolumn{2}{c}{0.1} \\
  $\eta$ & $\sim$1 & $\sim$4 \\ 
\end{tabularx}
\endgroup
\end{table}
\end{center}

% - - - - - - - - - - - - - - - - - - - - - - - - 
\subsection{Numerical model}
\label{sec:num_mod}
% - - - - - - - - - - - - - - - - - - - - - - - - 

We solve the equations presented in the previous section using the finite volume code \texttt{MPI-AMRVAC} \citep{Xia2017}. Our simulation box is a spherical grid centered on the donor star and extending from $1.2R_d$ up to 40 orbital separations, a distance at which ALMA can easily characterize the wind morphology for objects up to several 100 parsecs. The polar axis of the mesh is aligned with the orbital angular momentum vector and mirror-symmetry with respect to the orbital plane enables us to compute the solution only in the upper half of the spherical grid. Thanks to the geometric increase of the radial size of the cells, we can uniformly resolve the flow over a wide range of scales at an affordable computational cost \citep{ElMellah}. The coupling of this radially stretched grid with the block-based adaptive mesh refinement (AMR) available in the code enables us to increase the resolution where sharp density gradients appear, typically in the vicinity of the secondary object. Each level of AMR subdivides the block in 8 blocks, and each block has a size of 8$^3$ cells. On the coarsest AMR level, we work with 64 cells in the azimuthal direction, from 0 to 2$\pi$, and with 16 cells in the North-South direction, from 0 to $\pi/2$. The number of cells in the radial direction depends on the filling factor and mass ratios which set the position of the outer boundary of the simulation space, at 40 orbital separations. It typically ranges from 48 to 80 cells, which guarantees an approximate aspect ratio of one-to-one in the orbital plane.

The initial conditions are the ones for an isolated donor star derived from the 1D case (see Section\,\ref{sec:para_wind}). To determine the physical duration of each simulation, we compute the time required for a purely radial wind to cross the simulation box, from 1.2 dust condensation radii to 40 orbital separations, given the velocity profile and notwithstanding the presence of the secondary. The simulation duration is then set to 1.5 times this value which typically amounts to 4 to 20 orbital periods. Notice that a simulation duration as low as 4 is enough for the wind to cross the simulation space when the wind speed is much higher than the orbital speed. For winds with a lower terminal speed and a more progressive acceleration, a longer integration time is required to enable the wind to develop and reach the permanent regime. 

We work with a minimum number of AMR levels of 4 and the number of levels of refinement is set by the requirement that the accretion radius $R_{acc}$ is resolved with at least $\sim$10 cells along each direction. The accretion radius characterizes the extent of the bow shock formed ahead of the secondary as it gravitationally focuses the supersonic wind from the primary \citep{Edgar:2004ip}. Since it is the critical region where the subsequent disks and large scale structures originate, we need to ensure that we resolve it whatever the set of parameters we use. In the current problem, the accretion radius is given by:
\begin{equation}
    R_{acc}=\frac{2GM_2}{v^2_{\beta}\left(r=a\right)}
\end{equation}
and its normalized value with respect to the dust condensation radius depends only on the 4 first shape parameters given at the end of Section\,\ref{sec:bin_eq}. The maximum number of AMR levels needed, typically for a low mass of the secondary ($q=10$) and a large terminal wind speed ($\eta=8$), is prohibitively large, of the order of 12, and does not lead to significantly non-spherical circumbinary wind morphologies. Indeed, the faster the wind and the lighter the secondary, the more tenuous the effect of the presence of the secondary on the AGB wind morphology. Besides their expensive computational cost and the absence of large scale structures, these configurations do not lead to the formation of a wind-captured disk around the secondary due to the low amount of angular momentum in the wind when it reaches the secondary. These configurations yield very similar flow structures than the planar Bondi-Hoyle-Lyttleton problem \citep[][]{Blondin:2012vf,ElMellah2015}. Consequently, we lower by a factor of 10 the duration of the 13 simulations out of 72 which require strictly more than 6 levels of AMR. These simulations do not lead to the formation of visible patterns, except in the immediate vicinity of the secondary, well within its Roche lobe, where a tail appears but becomes quickly indistinguishable from the ambient essentially spherical wind. Among the remaining 59 simulations, the coupled use of the radially stretched grid with 4 to 6 levels of AMR leads to a wide dynamic range, with a ratio between the size of the lowest and highest resolution cell corresponding to 8 to 11 levels of AMR if we had used a Cartesian mesh.

In order to solve the aforementioned equations of hydrodynamics, we use a 3$^{\text{rd}}$ order HLL solver \citep{Toro1994} with a Koren slope limiter \citep{Vreugdenhil1993}. The equations have been discretized such that angular-momentum is preserved to machine precision, as described in \cite{ElMellah2018}. The gravitational softening radius is set by the size of the higher resolution cells. Doing so, we ensure that the wind-captured disk formed around the secondary is well resolved, at least along its radial extent and its outer regions.

% ------------------------------------------------
\section{Results}
\label{sec:results}
% ------------------------------------------------

How does the presence of a companion star or brown dwarf alter the structure of the stellar wind from the donor star? The influence of the secondary object is twofold. On one hand, its gravitational influence induces an orbital motion for the primary which itself can generate a spiral shock in the wind \citep{Kim2012}. And on the other hand, the gravitational pull of the secondary on the wind focuses it and leads to the formation of a detached bow shock around the secondary. Depending on the amount of specific kinetic energy deposited in the wind compared to the Roche potential, the flow structure around the secondary will either be essentially planar \citep{Blondin1991} or will embrace the more complex wind-RLOF configuration \citep{Mohamed2011}. Let us first examine the diversity of features which can emerge at large scale, in the circumbinary wind morphology.

% - - - - - - - - - - - - - - - - - - - - - - - - 
\subsection{Circumbinary wind morphology}
% - - - - - - - - - - - - - - - - - - - - - - - - 

\subsubsection{Equatorial density enhancement}
\label{sec:CEDE}

\begin{figure}
\centering
\includegraphics[width=0.98\columnwidth]{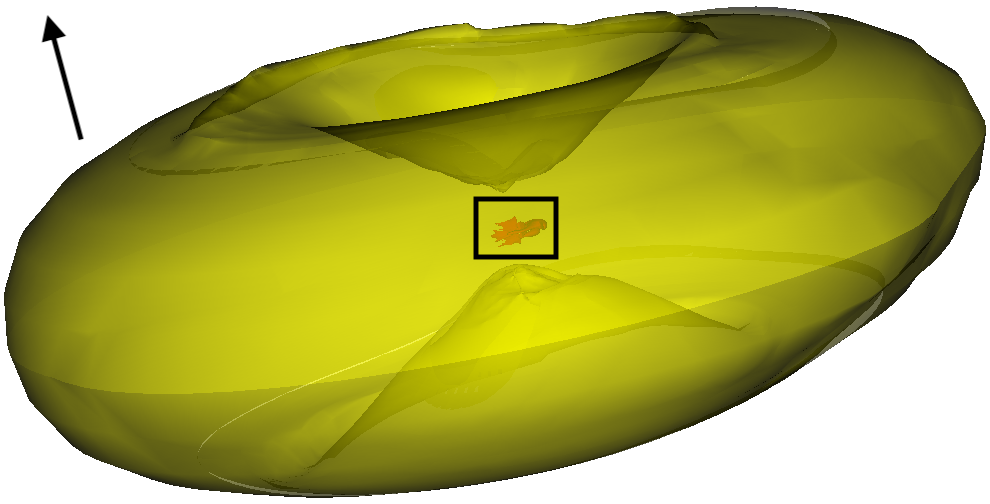}
\caption{3D iso-density contours of a simulation of a C-rich donor star with $q=1$, $f=80\%$ and $\eta=0.8$. If $R_d=6$ au and $\dot{M}_1=3\cdot$10$^{-5}$\msun yr$^{-1}$, the semi-transparent yellow surface surrounds a region where density is larger than 3$\cdot$10$^{-17}$g cm$^{-3}$ and with a diameter of approximately 800 au. The arrow indicates the direction of the orbital angular momentum. A zoom-in on the innermost region is provided in Figure\,\ref{fig:CEDE_zoom-in}.}
\label{fig:CEDE_large}
\end{figure} 

\begin{figure}
% 1col
\includegraphics[width=\columnwidth]{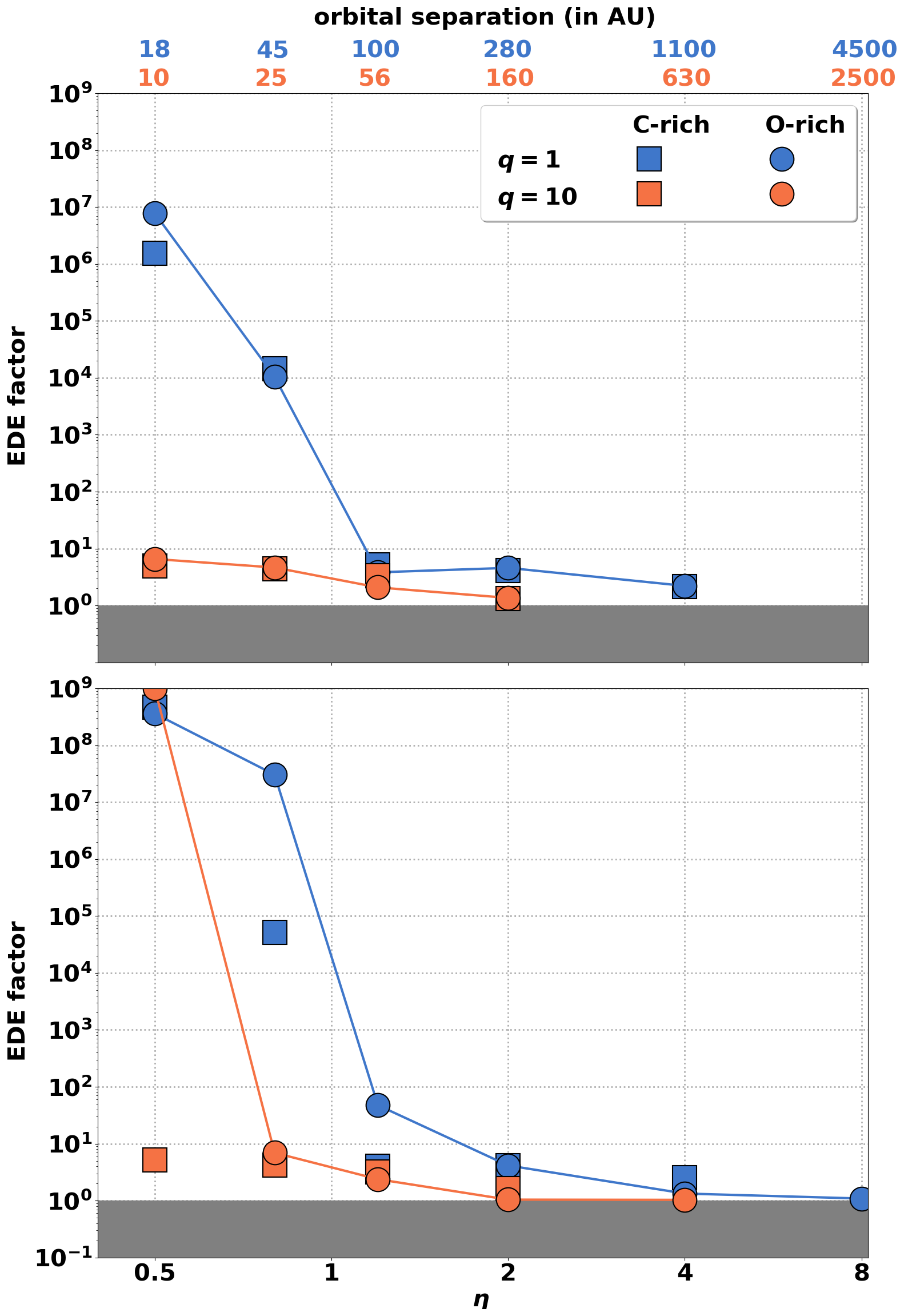}
\caption{EDE factors for each simulation as a function of $\eta$, the ratio of the terminal wind speed to the orbital speed, for a dust condensation radius filling factor of 5\% (top) and 80\% (bottom). The EDE factors as defined in the text evaluate how the wind is compressed in the orbital plane, with a value of 1 for a spherically symmetric flow. Different mass ratios (resp. $\beta$ exponents) are represented with different marker shapes (resp. colors). The solid lines connect the points corresponding to O-rich AGB donor stars. The top axis indicates the corresponding orbital separation for a donor mass of 4\msun, a terminal wind speed of 10\,km s$^{-1}$ (with $a\propto M_1/v_{\infty}^2$) and a mass ratio of 1 (in blue) and 10 (in orange).}
\label{fig:CEDE_factor}
\end{figure}

\begin{figure}
\includegraphics[width=\columnwidth]{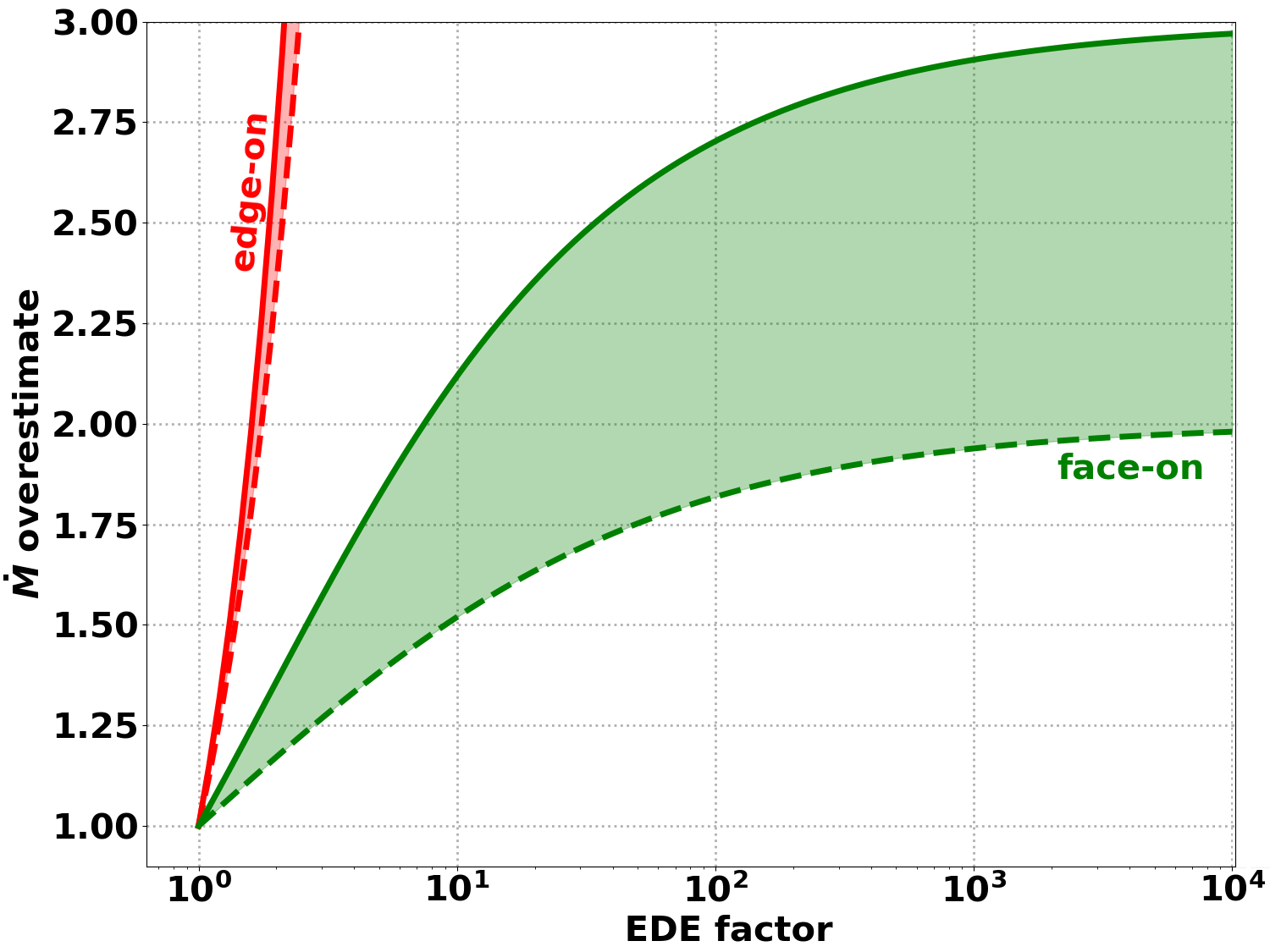}
\caption{Ratio of the mass-loss rate deduced from assuming spherical symmetry to the real mass-loss rate, as a function of the EDE factor for a system seen face-on (resp. edge-on) in green (resp. red). The solid and dashed lines are for an integration along a line-of-sight passing by the donor star and offset respectively.}
\label{fig:err_on_mdot}
\end{figure}

In the co-rotating frame where these simulations are ran, the orbital motion of the primary around the center of mass manifests itself through the centrifugal force. When the wind speed is low enough compared to the orbital speed (\ie for low values of the $\eta$ parameter), the wind is significantly beamed into the orbital plane. Without eccentricity, the flow remains essentially axisymmetric with respect to the donor star but a low density environment develops off-plane.

In Figure\,\ref{fig:CEDE_large}, three 3D iso-density contours have been represented at a scale extending up to 40 times the orbital separation (with a zoomed-in version in Figure\,\ref{fig:CEDE_zoom-in}). Most of the wind is concentrated in the orbital plane as shown by the semi-transparent yellow surface which represents an intermediate density. In red, within the black frame, the flow departs from axisymmetry due to the gravitational influence of the secondary. It is the scale of the orbital separation, at which wind-RLOF mass transfer takes place, and it is described in more detail in Section\,\ref{sec:innermost} where a zoomed-in figure can be found.

Recently, \cite{Decin2019} suggested that the measured mass-loss rates of O-rich intermediate mass AGB stars coined as OH/IR-stars could have been significantly overestimated because of the underlying assumption that the wind was spherically distributed. This preliminary estimate relied on ballistic and hydrodynamic simulations with prescribed acceleration profiles to reproduce different velocity laws \citep{Kim2012,ElMellah2016a}. Here, we introduce a factor to quantify the equatorial density enhancement (EDE) based on a subdivision of the flow in 3 regions each spanning a range of co-latitudes of $\pi/6$: 
\begin{itemize}
    \item \underline{region A:} within a co-latitude of $\pi/6$ around the axis normal to the orbital plane (the polar axis of our spherical mesh)
    \item \underline{region C:} within a co-latitude of $\pi/6$ around the orbital plane
    \item \underline{region B:} in-between the two
\end{itemize}
Once the flow has reached numerical equilibrium, we average the density in regions A and C (and over the $2\pi$ longitudinal angles). Then, for each radius, we divide the obtained mean density near the orbital plane by the one near the axis and take the median to obtain the EDE factor, represented in Figure\,\ref{fig:CEDE_factor}. In order to relate this factor to the overestimation of the mass-loss rate, we compute the ratio between the column density of an isotropic wind with a density equal to the one computed in the region C, and the column density of a wind subdivided in the 3 uniform regions described above. The results are reported in Figure\,\ref{fig:err_on_mdot} and depend on the inclination of the system with respect to the line-of-sight. The wind speed is assumed to be constant so as the density decays as $r^{-2}$ and then, when the environment is optically thin, the ratio of column densities is a good proxy for the ratio of the measured mass-loss rate assuming spherical symmetry to the real mass-loss rate. We perform this computation for systems seen face-on (in green) and edge-on (in red), and integrating along a line-of-sight either passing by the donor star or offset such as it intercepts the orbital plane (if face-on) or the orbital axis (if edge-on) with an impact parameter of $\sim$10 dust condensation radii. 

In Figure\,\ref{fig:err_on_mdot}, we retrieve that a EDE factor of unity is associated to a ratio of 1 (\ie no overestimation, the real value of the mass-loss rate is obtained). The error remains under a factor of 3 if the system is seen face-on. The error increases much faster with the EDE factor for systems seen edge-on (approximately proportionally). This computation shows that significant overestimates can occur even for EDE factors as low as a few. The detailed evaluation of the actual overestimation depends on the density profile assumed \citep{Homan2015,Homan2016} and on the diagnostics used (either molecular lines or dust emission). % Although this computation is based on a simplistic representation of the wind, it shows that significant overestimates can occur even for EDE factors as low as a few. The detailed evaluation of the actual overestimation depends on the density profile assumed.

\begin{figure*}[!b]
    \centering
    % 1col
    \includegraphics[width=1.95\columnwidth]{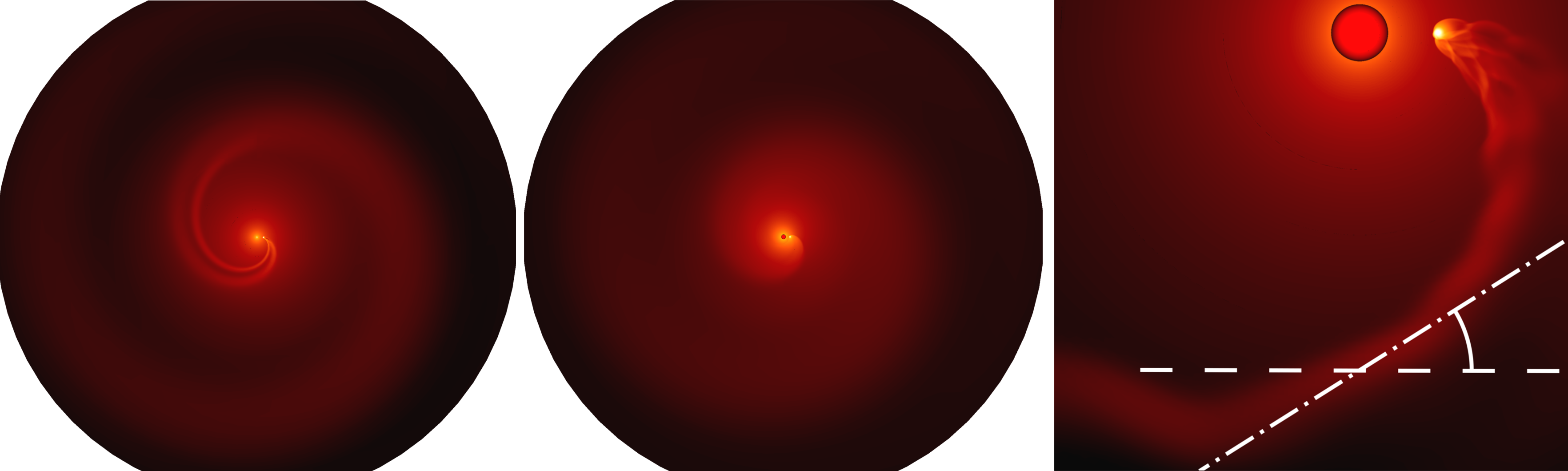}
    \caption{Logarithmic density maps in the orbital plane for 3 simulations with $q=1$ and $\eta=4$. (left panel) C-rich donor star with $f=5\%$. (center panel) O-rich donor star with $f=80\%$. (right panel) Zoom in on a C-rich donor star with $f=80\%$. Zoom in for the left and center panels and a zoom out for the righ panel are available as supplementary material online. The outer edge of the simulation space lies at 40 orbital separations away from the donor star.}
    \label{fig:spiral_3D}
\end{figure*}

As expected, for circumbinary envelopes essentially spherical, the EDE factor is close to 1. However, the EDE factor can be much larger when most of the mass is concentrated in the orbital plane. The latter case does occur for realistic parameters. The most direct conclusion we can draw from Figure\,\ref{fig:CEDE_factor} is the quick increase in the compression of the flow as soon as the terminal wind speed gets slightly smaller than the orbital speed. The exact position of the threshold in $\eta$ depends on the other parameters: the transition occurs at larger values of $\eta$ for a lower mass ratio and a larger filling factor. While the former evolution corresponds to a larger gravitational influence of the secondary, the latter yields less room for the wind to accelerate which makes the wind more prone to be shaped by the orbital motion. For a mass ratio of 10 and a filling factor of 5\%, the wind remains spherical with a compression factor below 10 for a terminal wind speed as low as half of the orbital speed (orange markers in top panel in Figure\,\ref{fig:CEDE_factor}). Notice however that such a low ratio $\eta$ can only be reached in extreme cases, when the orbital period is shorter than a few years \citep[for instance sequence E long period variables][]{Nicholls2010} or when the wind is anomalously slow.

A general trend to keep in mind is that once the terminal wind speed is slow enough compared to the orbital speed, the compression in the orbital plane is much more efficient for a dust condensation filling factor of 80\% (bottom panel) than 5\% (top panel). This is due to the fact that the wind speed at the orbital separation can be very low for large filling factors since the acceleration only starts shortly before the wind reaches the secondary. But once $\eta\gtrsim 2$, the compression of the flow is insensitive to the filling factor and to the chemical content of the dust. 

Furthermore, these results show that when the filling factor of the dust condensation radius is lower than 10\%, little difference should be expected between the compression of the wind around C-rich and O-rich AGB stars. Indeed, by the time it reached the secondary, the dust-driven wind had time to almost fully reach its terminal wind speed. In this configuration, provided a C-rich and an O-rich AGB stars have the same ratio $\eta$ of the terminal wind speed to the orbital speed, the morphology of their circumbinary envelope should be similar and no systematic difference should exist. Biases could only be due to different mass ratios between the donor star and the secondary companion. On the contrary, at low $\eta$ and for a dust condensation radius which extends up to 80\% of the Roche lobe radius of the primary (bottom panel in Figure\,\ref{fig:CEDE_factor}), the circumbinary envelope around an O-rich star is more prone to be compressed in the orbital plane than for a C-rich star due to the slow acceleration of O-rich winds.

Eventually, it must be noticed that up to a parameter $\eta$ of 2, a EDE factor of 3 to 8 can subsist when the secondary is massive enough ($q\lesssim10$) and/or when the filling factor is large enough. This implies that mass-loss rate estimates assuming a 1D geometry might significantly overestimate the actual mass-loss rate if the used diagnostic is sensitive to the equatorial density enhancement. As discussed by \cite{Decin2019}, mass-loss rates based on dust diagnostics are sensitive to the EDE, while low-excitation CO lines observed with large beams are more reliable tracers of the mass-loss rate even in the case of binary interaction. An additional concern is that the accretion of a fraction of the outflow by the secondary might lower the mass-loss rate measured at larger distance (up to $\sim$50\%, see Section 3.3), a contribution which impacts the long term orbital evolution.

% {DELETE: Depending on the density profile assumed in the wind to compute the mass-loss rate, it might still lead to significant overestimates. On the reverse, accretion of a fraction of the outflow by the secondary might lower the mass-loss rate measured at large distance, a contribution which impacts the long term orbital decay and that we treat in Section\,\ref{sec:ldot}.}

\subsubsection{Arcs \& spiral arms}

Spiral arms are a recurrent pattern around cool evolved stars such as the C-rich AGB stars R Sculptoris \citep{Maercker2012,Homan2015} and AFGL3068 \citep{Kim2017}. Arcs have also been observed as broken rings in protoplanetary nebulae by \cite{Ramos-Larios2016} for instance. They have been proposed to originate from expanding outflows collimated in the orbital plane and could be the manifestation of spiral arms seen edge-on \citep{Kim2019}.

The numerous 3D simulations we performed bring a unique opportunity to disentangle between the intrinsic characteristics and the apparent differences induced by the random inclination of the system with respect to the line-of-sight. In these simulations, the spiral shocks in the wind are always seeded in the vicinity of the secondary. However, at large scale, the details of the flow structure around the secondary do not impact the shape of the circumbinary envelope which is determined by the dimensionless parameters listed in Section\,\ref{sec:bin_eq}.

% It is why we retrieve similar large scale wind patterns as previous studies, including those where the bow shock in the vicinity of the secondary was not necessarily resolved. 

A spiral shock is defined by its density enhancement, pitch angle and vertical extension. The pitch angle is the angle between the spiral and the local direction of the circular orbit, with larger pitch angles for more radial spirals (see angle represented in the right panel in Figure\,\ref{fig:spiral_3D}). The larger the pitch angle, the larger the ratio of the inter-arm separation to the orbital separation. When the terminal wind speed is larger than the orbital speed, the pitch angle increases quickly with $\eta$ \citep[see][for a semi-analytic formula of the locus of the spiral]{Kim2012a} although the density contrast and vertical extent eventually vanish. In Figure\,\ref{fig:spiral_3D}, we represented slices in the orbital plane of the density of 3 simulations with a large terminal wind speed compared to the orbital speed, similar to what \cite{Cernicharo2015} found for CW Leo ($\eta=4$). Although they cannot be directly compared to the observations, they are indicative of what a face-on circumbinary envelope might look like in the zero-speed slice of a multi-channel molecular line emission map. A first striking result is the independence of the spiral locus, once scaled to the orbital separation, with the filling factor and the $\beta$ exponent. However, the more progressive acceleration of the O-rich AGB star (central panel) blurs the spiral shock which becomes less visible. On the other hand, a C-rich AGB donor star (left panel) can feature a sharp hollow spiral. In the right panel, we represented a zoom-in on the region where the shock first forms, for a simulation with a large filling factor C-rich donor star and a higher resolution than the two other panels. A strong instability sets in which propagates along the shock up to large scales. The stability of the Bondi-Hoyle-Lyttleton flow in 3D has been a long-debated question \citep[see][for an overview of the numerical simulations]{Foglizzo2005}. Different types of instabilities which can be triggered have been identified such as (i) the flip-flop instability, susceptible to produce alternatively co and counter-rotating wind-captured disks but seemingly absent from 3D simulations \citep{Ruffert1999,Blondin:2012vf}, and (ii) the advective-acoustic instability \citep{Foglizzo2006a}, extensively studied in spherical geometry. Here, we do not carry out the detailed analysis of the origin of this instability but notice its quick growth and its capacity to perturb the flow at large scale.

\begin{figure}
\centering
\includegraphics[width=0.99\columnwidth]{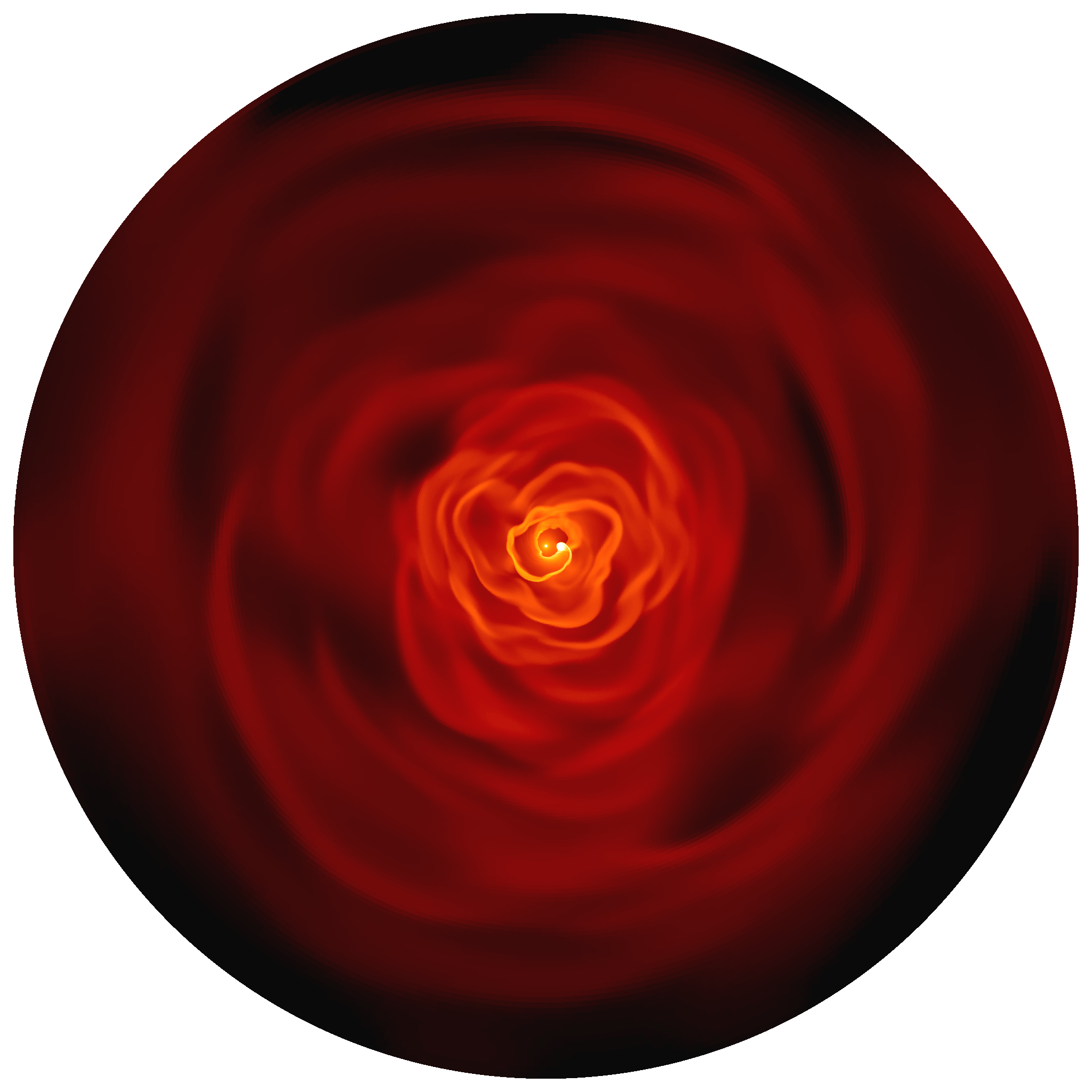}
\caption{Slice in the orbital plane of the gas density (logarithmic scale), after 20 orbital periods. This simulation is for a C-rich donor star with $q=1$, $f=20\%$ and $\eta=0.8$. An animated version can be found in the supplementary material.}
\label{fig:petals}
\end{figure}

When the terminal wind speed becomes lower than the orbital speed, the pitch angle significantly decreases as the spiral grows: the wind is not provided enough kinetic energy per unit mass by the stellar radiative field to overcome the attraction of the Roche potential ($\eta\lesssim 1$). A significant fraction of the wind fails to reach the outermost shells of the simulation box, located at 40 orbital separations (see also Section\,\ref{sec:ldot}). As visible in the high resolution simulation Figure\,\ref{fig:petals}, the spiral shock stalls, oscillates, breaks apart after a couple of orbital separations and forms a concentric petals pattern in the orbital plane, although the orbit is not eccentric and the stellar mass-loss rate is steady. At larger distances, it manifests as arcs in the orbital plane which propagates outwards in a self-similar way, which means that this morphology could be retrieved at larger distances than 40 orbital separations. At this stage, the propagation is made possible thanks to the pressure support of the inferior layers of the wind. We notice the striking similarity of the flow morphology in Figure\,\ref{fig:petals}, where $\eta=1$, with respect to the zero-speed slice in the molecular line emission maps of CW Leo by \cite{Cernicharo2015}. In contrast, using $\eta=4$ leads to the formation of a large pitch angle spiral similar to the right panel in Figure\,\ref{fig:spiral_3D}. This result leads us to bring support to the low orbital period derived by \cite{Decin2015} rather than the longer one proposed by \cite{Cernicharo2015}.

The stalling of the spiral in Figure\,\ref{fig:petals} does not form a proper circumbinary disk. In \cite{Chen2017}, their simulation "M1" with dimensionless parameters close to the one in Figure\,\ref{fig:petals} clearly led to the formation of a ring-like structure at a few orbital separations. We think this difference comes from our simplified treatment of the cooling, based on a polytropic prescription, which does not capture the enhanced cooling efficiency in the highest density regions, contrary to the more physically realistic cooling mechanism \cite{Chen2017} rely on. 

\begin{figure} %[!t]
    \centering
    \includegraphics[width=0.99\columnwidth]{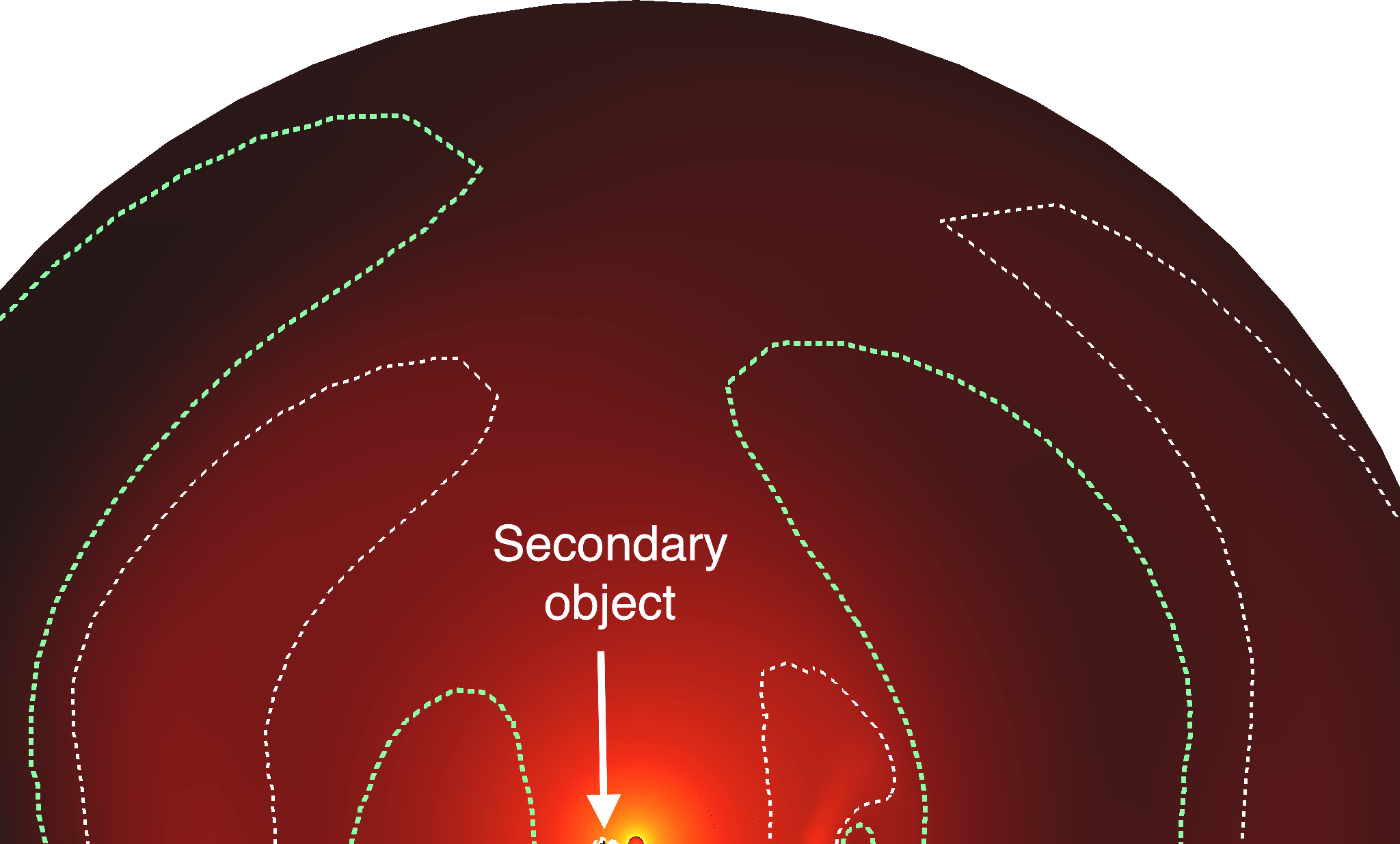}
    \caption{Edge-on view of the simulation in the right panel in Figure\,\ref{fig:spiral_3D}. The green and white dashed contours represent respectively 20\% deficit and excesses in the local mass-loss rate with respect to the isotropic case.}
    \label{fig:arcs_spir}
\end{figure}

The flow structure is largely different when seen edge-on. Since its compression in the orbital plane has been described in detail in Section\,\ref{sec:CEDE}, let us focus here on the arcs which emerge as the $\eta$ parameter decreases. In Figure\,\ref{fig:arcs_spir}, we show the density map of the same simulation as the right panel in Figure\,\ref{fig:spiral_3D} but seen from another viewpoint. It is a slice containing the orbital angular momentum vector and the axis joining the two bodies, with the secondary lying on the left of the primary. With a EDE factor of $\sim3$, the wind compression in the orbital plane is hardly visible by eye in the density distribution. However, the monitoring of the local mass-loss rate (\ie per unit solid angle) and its comparison with the isotropic one reveals alternating excesses and deficit (respectively white and green dashed lines) of more than 20\%. They are essentially due to an enhanced density and have opposite phase on the two sides of the secondary. If non-linear effects are triggered in these regions such as dust condensation \citep{Boulangier2019}, these regions might still be visible for an $\eta$ parameter as large as 4. Otherwise, we have to turn towards lower values of $\eta$ such as in the bottom panel in Figure\,\ref{fig:wind-RLOF}. It represents only the inner region but arcs appear in the flow as slices of a spiral propagating outwards (see top panel in the same figure) and spreads to much larger distances.

% - - - - - - - - - - - - - - - - - - - - - - - - 
\subsection{Innermost region: wind-captured disks}
\label{sec:innermost}
% - - - - - - - - - - - - - - - - - - - - - - - - 

Roche lobe overflow is a well-known mass transfer mechanism which leads to the formation of a narrow stream of material flowing through the inner Lagrangian point. As the material falls towards the secondary object, it acquires enough angular momentum to form a large and permanent disk within its Roche lobe \citep[see \eg][]{Frank2002}. Here, we look at a somewhat less extreme case since most cool evolved stars with undetected periodic velocity modulation along the line-of-sight are expected to have an orbital separation large enough to currently not fill their Roche lobe. At the other end of the possible mass transfer mechanisms lies wind accretion, based on the theory of planar accretion of a supersonic flow by a point-mass \citep[section 8.1 in][]{ElMellah2016}. In our simulations, this regime is an accurate representation of the configurations when the wind speed is much higher than the orbital speed when the wind reaches the secondary (\ie low $f$ and high $\eta$). But for wind speeds of the order of the orbital velocity, the wind is strongly beamed towards the secondary and we find a third type of mass transfer called wind-RLOF \citep{Mohamed2007}. Due to the large stellar mass-loss rate and to the focusing of the wind in the orbital plane, a significant mass transfer enhancement can take place compared to the two other regimes. It has been suggested to explain the origin of the excess in barium of some carbon-enhanced metal-poor stars \citep{Abate2013} but also for the high mass accretion rates needed to reproduce the luminosity of some ultra-luminous X-ray sources \citep{ElMellah2018a}.

%  {??( LEEN: REF? SYSTEMATIC MEASURE OF Vr?)}

\begin{figure}
\centering
\includegraphics[width=0.99\columnwidth]{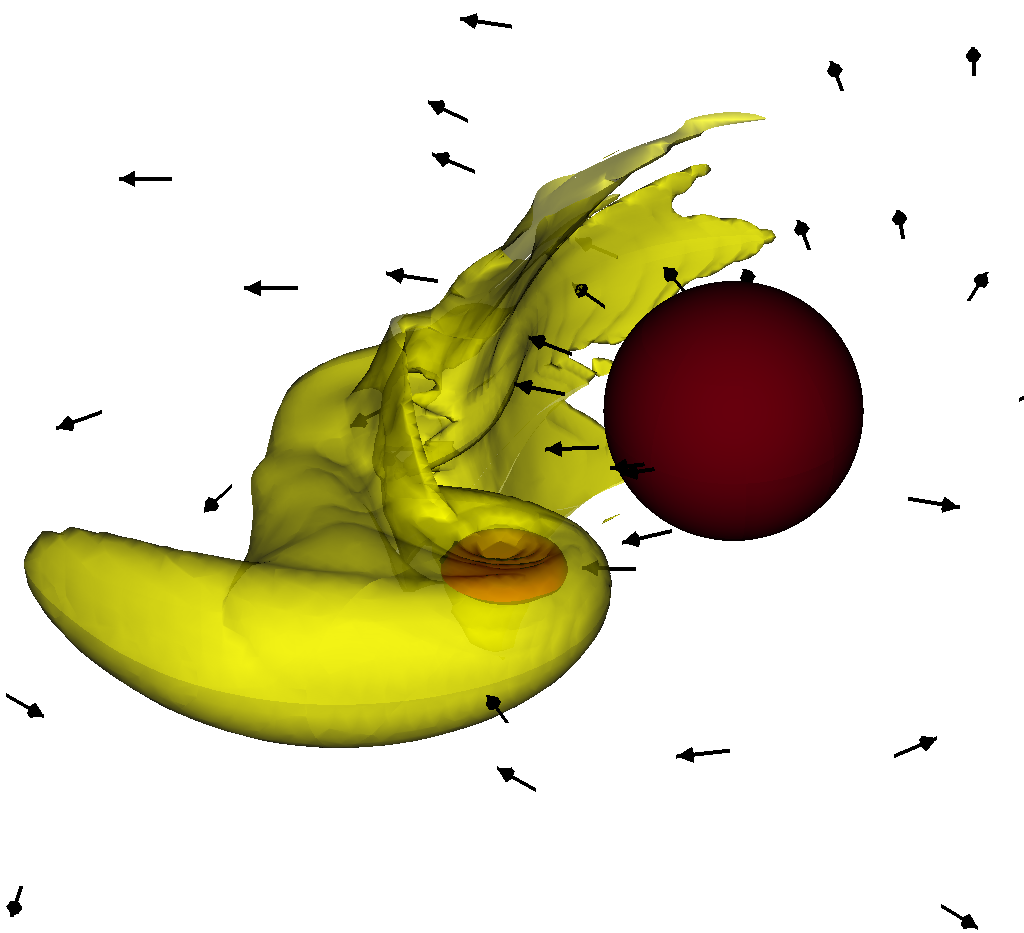}
\caption{Zoom in on the central region in Figure\,\ref{fig:CEDE_large} (black frame) where the donor is a C-rich star and $q=1$, $f=80\%$ and $\eta=0.8$. If $R_d=6$ au and $\dot{M}_1=3$10$^{-5}$\msun yr$^{-1}$, in semi-transparent yellow is represented the 3D iso-density surface $\rho=9\cdot$10$^{-14}$g cm$^{-3}$ and in red, the surface $\rho=9\cdot$10$^{-12}$g cm$^{-3}$. The inner boundary of the simulation space is visible in dark red in the right part and the velocity field in the orbital plane has been represented.}
\label{fig:CEDE_zoom-in}
\end{figure}

\begin{figure*}
  \centering
  % 1col
 \includegraphics[width=1.73\columnwidth]{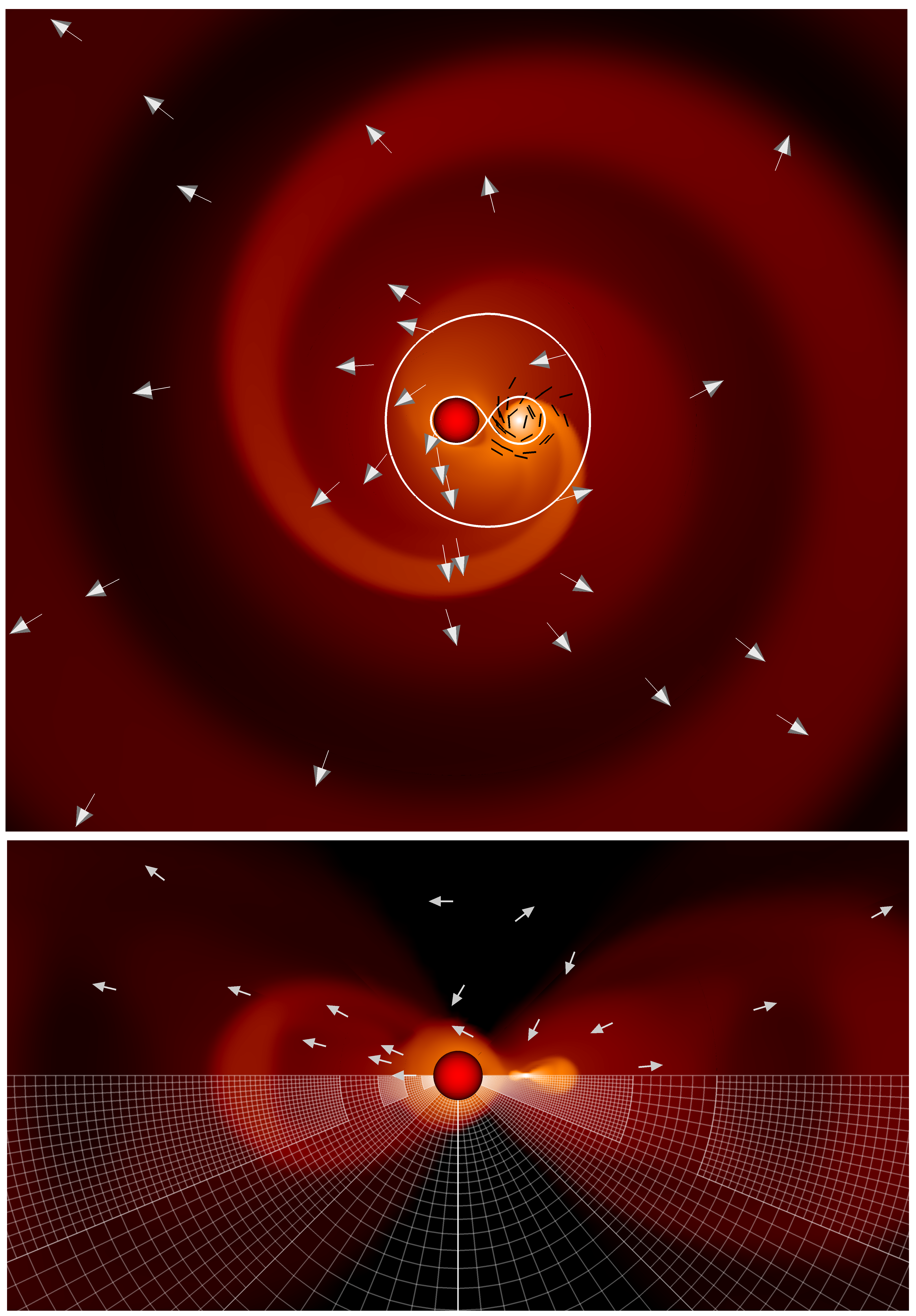}
\caption{(top panel) Logarithmic density map in the orbital plane of the innermost region of the wind. The arrows in white represent the velocity field while the black lines indicate the streamlines in the vicinity of the secondary where a wind-captured disk develops. In solid white is the contour of the Roche potential with the value at the innermost Lagrangian point. (bottom panel) Side-view of the same density map, with the secondary object on the right of the inner boundary (the red sphere of radius 1.2 dust condensation radius). The velocity field has been represented in the upper half while the mesh and its different levels of refinement are visible in the bottom half. This simulation is for an O-rich donor star with $q=1$, $f=80\%$ and $\eta=0.8$.}
\label{fig:wind-RLOF}
\end{figure*}

Our simulations reproduce the main features associated to the wind-RLOF mechanism: the enhanced mass transfer and the formation of a wind-captured disk around the accretor. In Figure\,\ref{fig:wind-RLOF}, we show how the wind is strongly shaped by the gravitational potential of the two orbiting bodies in the innermost regions of a wind emitted by an O-rich AGB star. In this configuration, the wind is so slow when it reaches the secondary that the accretion radius is of the order of the Roche lobe radius of the secondary, a clear signature of wind-RLOF. The top panel is a slice in the orbital plane, associated to a face-on view, while the bottom panel represents a transverse slice containing the secondary object (on the right of the primary). The velocity field, like in all figures in this paper, is the one in the inertial observer frame. While it remains essentially radial around the donor star beyond a few orbital separations, a high vorticity indicative of a wind-captured disk arises in the vicinity of the secondary, on the right of the AGB donor in the top panel in Figure\,\ref{fig:wind-RLOF}. Notice the differences between the flow structure in the top panel of this figure or in the zoom in 3D representation in Figure\,\ref{fig:CEDE_zoom-in}, and in the right panel in Figure\,\ref{fig:spiral_3D} where the wind was much faster with respect to the orbital speed: for the slower wind in Figures\,\ref{fig:CEDE_zoom-in} and \ref{fig:wind-RLOF}, the accretion radius gets larger than the Roche lobe radius. The flow gains enough angular momentum to form a permanent wind-captured disk of maximal extension \citep{Paczynski1977} and a wide spiral with a low pitch angle in the wake of the accretor.

Furthermore, the wind is now strongly beamed into the orbital plane as visible in the bottom panel in Figure\,\ref{fig:wind-RLOF}. Off the plane, the wind fails to take off and falls back, leading to a significant increase in the density scale compared to the configurations where the wind remains essentially isotropic (\eg Figure\,\ref{fig:arcs_spir}). Within this compressed flow, the spiral visible in the upper panel deploys and seen edge-on, it manifests with alternating arcs of higher and lower density. 

The presence or not of a wind-captured disk around the secondary shares another connection with observations. Indeed, provided the secondary has a magnetic field large enough, the accretion of matter might not only change the spin of the secondary but also lead to significant outflows \citep{Zanni2013}. Depending on how collimated they are and on the inclination of the magnetic field with respect to the orbital angular momentum axis, it could participate in the shaping of the low density off-plane environment.

% - - - - - - - - - - - - - - - - - - - - - - - - 
\subsection{Mass and angular momentum loss}
\label{sec:ldot}
% - - - - - - - - - - - - - - - - - - - - - - - - 

\begin{figure}
\centering
\includegraphics[width=0.99\columnwidth]{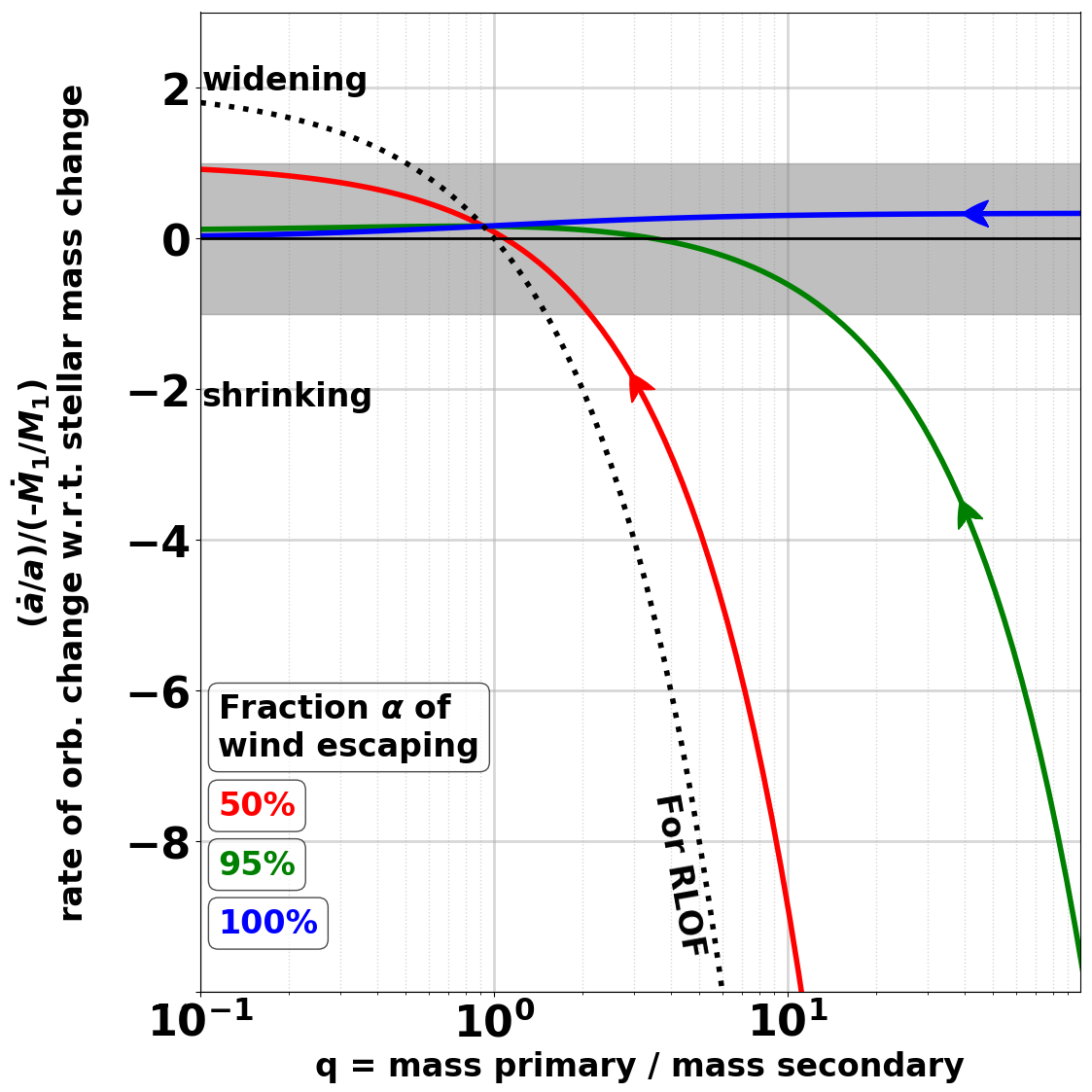}
\caption{Rate of orbital separation change compared to the rate of stellar mass change. In the grey shaded region, the mass of the donor star changes faster than the orbital separation while above (resp. below) the orbit expands (resp. contracts). The two limit cases are conservative mass transfer (dotted line, RLOF) and pure mass-loss without accretion by the secondary (solid blue line). In-between, the green and red solid lines are for an accretion efficiency by the secondary of 5\% and 50\% respectively. The arrows indicate that as mass transfer proceeds, the mass ratio can only decrease.}
\label{fig:orb_decay}
\end{figure}

In this part, we report on the fraction of wind which eventually manages to escape the binary system. Due to the preliminary expansion of the donor star, we neglect its spin compared to the much larger orbital angular momentum but \cite{Chen2018} carried out a detailed analysis of the orbital shrinking/widening where they also account for spin-orbit synchronisation. The secular change of orbital angular momentum induced by the mass-loss of the primary can only provoke a widening or a shrinking of the orbit, sometimes at a rate much higher than the mass-loss itself. Constraining the angular momentum loss is a necessary step to perform binary population synthesis and evaluate which fraction of the systems will undergo a common envelope phase \citep{Ricker2012} or produce a Type Ia supernova \citep{IbenI.1984,Webbink1984}. Here, given the large stellar mass-loss rate and the large orbital separation, additional sources of orbital angular momentum loss or gain (\eg magnetic braking, gravitational waves and stellar expansion) are believed to be minor. If we assume that all the wind either leaves the system or is accreted by the secondary, we have the following evolution of the orbital period $P$ \citep{Tauris2003}:
\begin{equation}
\begin{aligned}
    \frac{\dot{P}}{P}&=\frac{\dot{M}_1}{M_1}\left[\frac{3\alpha}{1+q}+\frac{q\left(1-\epsilon\right)}{1+q}-3\left(1-\epsilon q\right)\right]\\
    &=\frac{\dot{M}_1}{M_1}\left[\frac{3q^2\left(1-\alpha\right)-2\alpha q-3\left(1-\alpha\right)}{1+q}\right]
\end{aligned}
\end{equation}
where $\alpha$ is the fraction of the wind which escapes the system while $\epsilon$ is the fraction of wind captured by the secondary (\aka the accretion efficiency). The last expression is obtained using $\alpha+\epsilon=1$. Since $\dot{M}_1$ is always negative (the donor star looses mass), the sign of the term within brackets decides whether the orbit shrinks (if positive) or widens (if negative). In Figure\,\ref{fig:orb_decay}, we represented the opposite of the term within bracket which, sign apart, stands for the ratio of two characteristic time scales: the amount of time for the star to loose a given fraction of its mass to the amount of time for the orbital separation to be modified by the same fraction. In Appendix\,\ref{app:app3} and for comparison with other numerical results, we provide the reader with the same graphics for the orbital period and the orbital speed. 

The mass-loss rate $\dot{M}_1$ is highly variable during the evolution of AGB stars, with short episodes of enhanced mass-loss, up to 10$^{-5}$\msun yr$^{-1}$, separated by quiescent phases where the mass-loss rate is of the order of 10$^{-7}$\msun yr$^{-1}$ \citep{Bloecker1995}. In this context, the reader should keep in mind that the time scales involved in the ratio represented in Figure\,\ref{fig:orb_decay} are bound to change by order of magnitudes during the AGB phase and should be thought as instantaneous rates of change. Within the grey shaded region, the rate of change of the orbital separation is lower than the rate of change of the mass of the donor star while elsewhere, for a given relative change of the donor mass by a certain fraction, the orbital separation changes by more than this fraction (\ie the orbital separation evolves "faster" than the donor mass). Not surprisingly, for a mass ratio $q=M_1/M_2$ lower than unity, the orbital separation evolves slowly. Indeed, since most of the mass is now in the secondary, the total mass remains fairly constant as the primary looses mass and the impact of the mass-loss on the orbit is limited.

% For a 1\msun star loosing mass at a rate of $10^{-5}$\msun yr$^{-1}$, the mass-loss time scale is 10$^5$ years ( LEEN: COMMENTS ON HOW STABLE THE MASS LOSS RATE IS OVER THIS DURATION? {Incorrect statement, since then the star would weigh 0\msun. A start with initial mass of 1\msun\ starts its post-AGB phase when the core mass is around 0.5\msun ; see Karakas2014. These stars are not expected to have an Mdot of 1e-5msun/yr for a long time, even seldomly they will have a mass-loss rate above a few times 1e-7msun/yr. To be discussed}).

The two asymptotic cases are the black dotted line (RLOF, $\epsilon=1$) and the solid blue line (free wind, $\alpha=1$). Arguments based on a polytropic representation of the equation-of-state of the donor star show that for $q\sim 1$, a purely conservative mass transfer (\ie $\epsilon=1$ so $\alpha=0$) is prone to provoke a faster contraction of the Roche lobe radius than of the stellar radius \citep{DSouza:2005jx}. The positive feedback loop provokes the quick inspiral of the two bodies and the system enters a common envelope phase, although properly accounting for the stellar internal structure pushes the critical mass ratio to a few \citep{Pavlovskii2017,Quast2019a}. These results are consistent with the present framework when mass transfer is only partly conservative ($\alpha<1$, black dotted line, red and green solid lines in Figure\,\ref{fig:orb_decay}): we retrieve in Figure\,\ref{fig:orb_decay} that there is always a critical mass ratio beyond which the mass transfer yields a quick inspiral of the system (compared to the mass-loss time scale). The less conservative the mass transfer, the larger this critical mass ratio. 

\begin{figure}
% 1col
\includegraphics[width=\columnwidth]{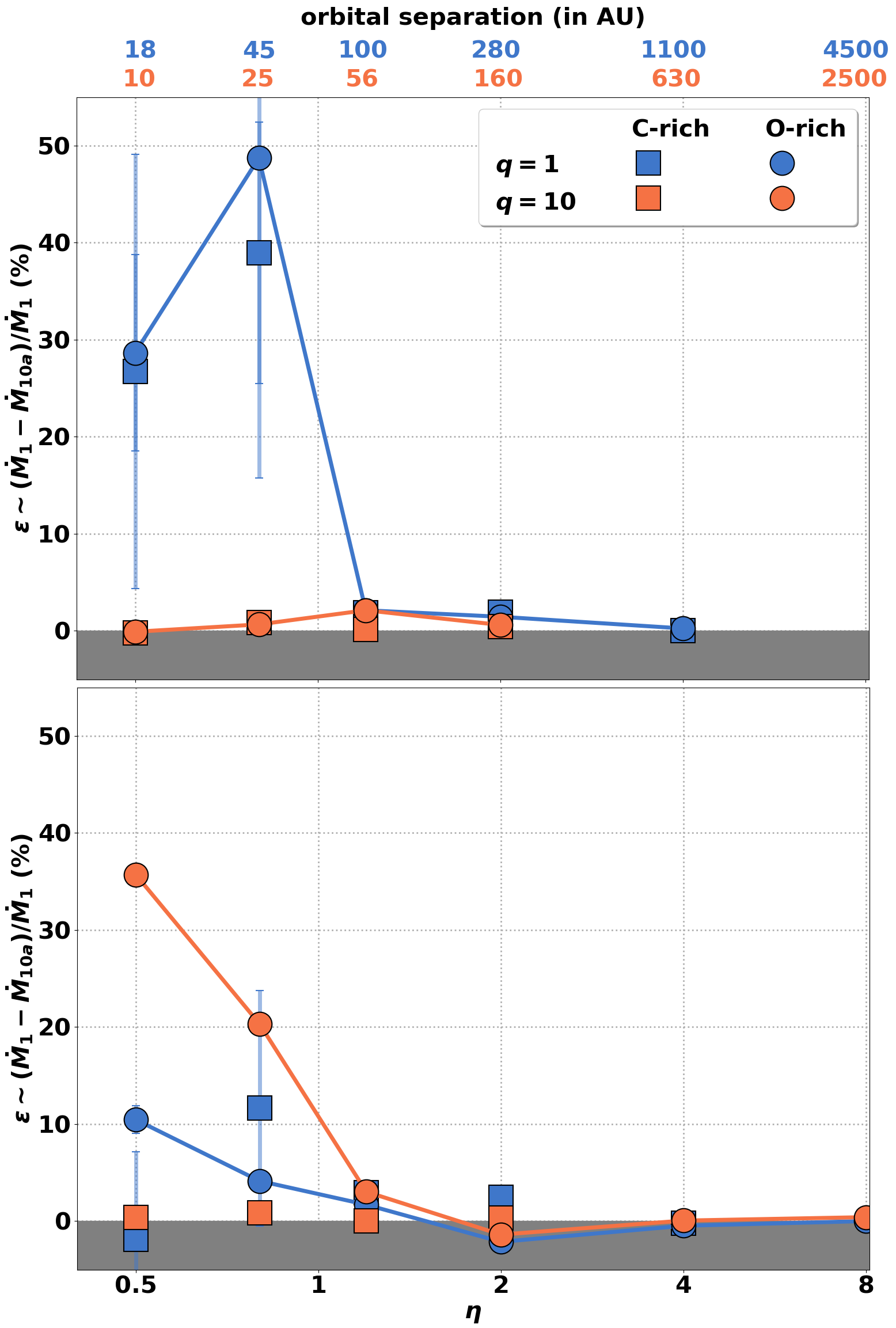}
\caption{Fraction $\epsilon$ of the wind captured by the secondary (accretion efficiency) as a function of the ratio $\eta$ of the terminal wind speed to the orbital speed for a dust condensation radius filling factor of 5\% (upper panel) and 80\% (lower panel). Square (resp. circular) markers are for C-rich (resp. O-rich) AGB stars. Blue (resp. orange) markers are for mass ratios $q=1$ (resp. $q=10$). The results for O-rich stars are connected with solid lines and the error bars represent the variations over a dynamical time scale. The top scale is the same as in Figure\,\ref{fig:CEDE_factor}.}
\label{fig:alpha}
\end{figure}

We measure $\alpha$ by computing the mass-loss rate at 10 orbital separations and comparing it to the mass-loss rate at the inner edge of the simulation space, representative of the stellar mass-loss rate. In Figure\,\ref{fig:alpha}, we plotted $\epsilon=1-\alpha$ as a function of the main parameter, the ratio $\eta$ of the terminal wind speed to the orbital speed. Beyond $\eta=1$, the accretion efficiency remains invariably smaller than 5\%. Negative values are due to unsteadiness and uncertainties in the numerical computation, of the order of a few percent. This uncertainty precludes any quantitative prediction, but relying on Figure\,\ref{fig:orb_decay} provides insightful predictions. According to these results, for a mass ratio inferior to 10, a terminal wind speed larger than the orbital speed means that the orbital separation remains essentially unchanged over the mass-loss time scale $M_1/\dot{M}_1$ (although the orbital period might increase significantly, see upper panel in Figure\,\ref{fig:orb_sep_decay}). In the same way, even for $\eta<1$, the orbital evolution of binaries where the donor star is a C-rich AGB stars ten times more massive than the secondary (orange squares) can only be modest, and so is the orbital evolution of O-rich AGB stars with $q=10$ provided the dust condensation filling factor is smaller than $\sim20$\% (orange circles in the upper panel). For $f=80$\%, the wind from O-rich AGB stars is significantly captured by the secondary when $\eta<1$ due to the wind-RLOF regime, leading to accretion efficiency of several 10\%. Provided the mass ratio is large ($q$ larger than a few), the orbital separation quickly decays until the mass ratio is low enough to enter the gray shaded region in Figure\,\ref{fig:orb_decay}. The associated quick increase of the orbital speed, still for $q$ larger than a few, will enhance the phenomenon by lowering $\eta$ and increasing the accretion efficiency $\epsilon$. In the same way, even for a low filling factor, a massive secondary object ($q=1$) will capture a significant fraction of the stellar wind but due to the low mass ratio, Figure\,\ref{fig:orb_decay} indicates that the orbit will only marginally change over a mass-loss characteristic time scale.  

% For some configurations with a large filling factor and a low terminal speed, part of the wind falls back onto the inner boundary of the simulation space: due to tidal effects, the effective mass-loss rate is lower than the theoretical isotropic mass-loss rate at the sonic point. It is this effect which makes the effective mass-loss rate lower for lower $q$, at low $\eta$ and for an O-rich star, which yields an apparent ratio of matter trapped in the system larger for larger $q$. In this regime, the absolute values of this ratio should not be taken literally.

% ------------------------------------------------
\section{Discussion \& summary}
\label{sec:discussion}
% ------------------------------------------------

With our numerical setup, we recover a wide variety of observed circumstellar morphologies produced by the presence of a lower mass companion around a mass loosing cool evolved star. We show that the main parameter driving the wind dynamics is the ratio $\eta$ of the terminal wind speed to the orbital speed, although significant differences can arise between C-rich and O-rich stars for $\eta\sim1$ when the dust condensation region extends up to 80\% of the Roche lobe radius of the donor star. These numerical simulations confirm that the observation of alternating higher and lower density arcs or of a spiral pattern are indicative of respectively an edge-on or a face-on inclination of the orbit with respect to the line-of-sight. We also computed the amplitude of the wind compression in the orbital plane for each set of parameters. It showed that beyond a mass ratio of 10, the wobbling motion of the primary induced by the secondary is not large enough to significantly flatten, except for a large dust condensation filling factor of 80\% and a very low ratio $\eta$ of 0.5 realistic only if the orbital period is lower than a decade. Assuming the orbital angular momentum changes only due to the outflow from the cool evolved star, we identified three necessary conditions to observe a significant shrinking of the orbit over a mass-loss time scale:
\begin{itemize}
    \item a terminal wind speed smaller than the orbital speed
    \item a mass ratio larger than a few
    \item a dust condensation region which extends up to several 10\% of the primary Roche lobe radius
\end{itemize}
In the other cases, the system will undergo a variation of the orbital separation of lower relative change than the stellar mass change, either an orbit widening or shrinking depending on the mass ratio and on the fraction of stellar wind captured by the secondary. Accounting for eccentricity will likely modify these conclusions. Eccentric orbits have been commonly observed for systems with large orbital separations (\ie large values of $\eta$), in agreement with the long circularization time scales expected \citep{Zahn1977}. Its effect on the wind morphology has been thoroughly reported on in \cite{Kim2015a}, \cite{Kim2017} and \cite{Kim2019} but to our knowledge, its impact on the orbital angular momentum remains to be investigated. 

In the configurations where $q\sim 1$ and the orbital speed is similar or larger than the terminal wind speed ($\eta\lesssim 1$), we observed the emergence of an unstable flow where the pitch angle of the spiral cancels within a couple of orbital separations. The shell fragments and is eventually ejected by the incoming material, producing a concentric petals pattern in the orbital plane reminiscent of what is observed around the C-rich AGB star CW Leo \citep{Decin2015,Cernicharo2015}. Most material is concentrated in the orbital plane due to the low value of $\eta$ so a slice in the orbital plane of the simulation should be representative of a face-on view. Interestingly enough, these successive arcs in the orbital plane appear in spite of the steadiness of the mass-loss rate, without any burst nor pulsation. The spacing between them is of the order of a few orbital separations at the outer edge of our simulation box and expands self-similarly with the distance to the donor star in the outer regions.

Unfortunately, the long orbital periods of binaries containing a cool evolved star preclude any monitoring of the rate of change of the orbital period. But in systems with shorter orbital periods such as supergiant X-ray binaries and ultra-luminous X-ray sources, this rate has been measured \citep[see \eg][]{Falanga2015}. The mass ratio can then be constrained using the wobbling motion of the two bodies projected along the line-of-sight \citep[see \eg][]{Quaintrell2003} or the timing of the neutron star pulses \citep[][]{Fuerst2018}. Together, they bring strong constraints on the wind fraction $\epsilon$ captured by the secondary and can be confronted to the mass accretion rates deduced from the X-ray luminosity.

% it must be noticed that red supergiant candidates have been found in orbit with accreting compact objects in 3 ultra-luminous X-ray sources \citep{Heida2015,Heida2016}. 

This paper introduces a convenient framework to implement new effects. In particular, the spherical mesh centered on the donor star opens the door to the addition of a key-ingredient: matter-radiation interaction. In this paper, we simply parametrized the wind acceleration to identify archetypal configurations. A subset of configurations can now serve as the bedrock for more realistic simulations where additional physical effects are introduced such as the growth of dust grains and their coupling with the stellar radiative field. Thanks to the low computational cost of each simulation and to innovative schemes to treat parabolic and elliptic equations recently implemented in \texttt{MPI-AMRVAC} \citep{Teunissen2019}, we can start to explore radiative effects.

% reasonably think that more physically demanding simulations will remain affordable with the present numerical setup. 

Radiation is not only a decisive actor in the launching and subsequent carving of the wind but also our main source of information on these systems. The advent of high-resolution imagery in (sub)millimeter wavelengths brought up the pressing need to produce synthetic observables to be confronted with. Each of the simulations we performed, which contain the information on the temperature, gas density and velocity 3D distribution, can now be post-processed to produce dust infrared and molecular-line emission maps for different inclinations, either with Monte-Carlo radiative transfer code such as SKIRT \citep{Baes2011,Camps2015}, RADMC-3D \citep{Dullemond2012} or SPARX \citep{Kim2013}, but also with new-generation ray solvers such as Magritte \citep{Ceuster2019a}.

\begin{acknowledgements}
The authors wish to thank the {\sc{atomium}} ALMA Large Programme collaboration (2018.1.00659, PI. L. Decin) for the observational consequences of the present analysis. IEM has received funding from the Research Foundation Flanders (FWO) and the European Union's Horizon 2020 research and innovation program under the Marie Sk\l odowska-Curie grant agreement No 665501. LD, JB and WD acknowledge support from the ERC consolidator grant 646758 AEROSOL. The simulations were conducted on the Tier-1 VSC (Flemish Supercomputer Center funded by Hercules foundation and Flemish government). RK is supported by Internal Funds KU Leuven, project C14/19/089 TRACESpace.
% THANK REFERREE
\end{acknowledgements}

%-------------------------------------------------------------------

\bibliographystyle{aa} 
\begin{tiny}
\bibliography{aa}

\begin{thebibliography}{104}
\expandafter\ifx\csname natexlab\endcsname\relax\def\natexlab#1{#1}\fi

\bibitem[{Abate {et~al.}(2013)Abate, Pols, Izzard, Mohamed, \&
  de~Mink}]{Abate2013}
Abate, C., Pols, O.~R., Izzard, R.~G., Mohamed, S.~S., \& de~Mink, S.~E. 2013,
  Astron. Astrophys., 552, A26

\bibitem[{Baes {et~al.}(2011)Baes, Verstappen, {De Looze}, Fritz, Saftly,
  {Vidal P{\'{e}}rez}, Stalevski, \& Valcke}]{Baes2011}
Baes, M., Verstappen, J., {De Looze}, I., {et~al.} 2011, Astrophys. Journal,
  Suppl. Ser., 196

\bibitem[{Bidelman \& Keenan(1951)}]{Bidelman1951}
Bidelman, W.~P. \& Keenan, P.~C. 1951, Astrophys. J., 114, 473

\bibitem[{Bloecker(1995)}]{Bloecker1995}
Bloecker, T. 1995, Astron. Astrophys. -Berlin-, 297, 727

\bibitem[{Blondin \& Raymer(2012)}]{Blondin:2012vf}
Blondin, J.~M. \& Raymer, E. 2012, Astrophys. J., 752, 30

\bibitem[{Blondin {et~al.}(1991)Blondin, Stevens, \& Kallman}]{Blondin1991}
Blondin, J.~M., Stevens, I.~R., \& Kallman, T.~R. 1991, Astrophys. J., 371, 684

\bibitem[{Boulangier {et~al.}(2019{\natexlab{a}})Boulangier, Clementel, {Van
  Marle}, Decin, \& {De Koter}}]{Boulangier2018}
Boulangier, J., Clementel, N., {Van Marle}, A.~J., Decin, L., \& {De Koter}, A.
  2019{\natexlab{a}}, Mon. Not. R. Astron. Soc., 482, 5052

\bibitem[{Boulangier {et~al.}(2019{\natexlab{b}})Boulangier, Gobrecht, Decin,
  de~Koter, \& Yates}]{Boulangier2019}
Boulangier, J., Gobrecht, D., Decin, L., de~Koter, A., \& Yates, J.
  2019{\natexlab{b}}, Mon. Not. R. Astron. Soc., 489, 4890

\bibitem[{Bujarrabal {et~al.}(2016)Bujarrabal, Castro-Carrizo, Alcolea,
  Santander-Garc{\'{i}}a, {Van Winckel}, \& {S{\'{a}}nchez
  Contreras}}]{Bujarrabal2016}
Bujarrabal, V., Castro-Carrizo, A., Alcolea, J., {et~al.} 2016, Astron.
  Astrophys., 593, A92

\bibitem[{Camps \& Baes(2015)}]{Camps2015}
Camps, P. \& Baes, M. 2015, Astron. Comput., 9, 20

\bibitem[{Castor {et~al.}(1975)Castor, Abbott, \& Klein}]{Castor1975}
Castor, J.~I., Abbott, D.~C., \& Klein, R.~I. 1975, Astrophys. J., 195, 157

\bibitem[{Cernicharo {et~al.}(2015)Cernicharo, Marcelino, Ag{\'{u}}ndez, \&
  Gu{\'{e}}lin}]{Cernicharo2015}
Cernicharo, J., Marcelino, N., Ag{\'{u}}ndez, M., \& Gu{\'{e}}lin, M. 2015,
  Astron. Astrophys., 575, 1

\bibitem[{Chen {et~al.}(2018)Chen, Blackman, Nordhaus, Frank, \&
  Carroll-Nellenback}]{Chen2018}
Chen, Z., Blackman, E.~G., Nordhaus, J., Frank, A., \& Carroll-Nellenback, J.
  2018, Mon. Not. R. Astron. Soc., 473, 747

\bibitem[{Chen {et~al.}(2017)Chen, Frank, Blackman, Nordhaus, \&
  Carroll-Nellenback}]{Chen2017}
Chen, Z., Frank, A., Blackman, E.~G., Nordhaus, J., \& Carroll-Nellenback, J.
  2017, Mon. Not. R. Astron. Soc., 468, 4465

\bibitem[{Chiţǎ {et~al.}(2008)Chiţǎ, Langer, {Van Marle},
  Garc{\'{i}}a-Segura, \& Heger}]{Chita2008}
Chiţǎ, S.~M., Langer, N., {Van Marle}, J., Garc{\'{i}}a-Segura, G., \& Heger,
  A. 2008, Astron. Astrophys., 488

\bibitem[{Claeys {et~al.}(2014)Claeys, Pols, Izzard, Vink, \&
  Verbunt}]{Claeys2014}
Claeys, J.~S., Pols, O.~R., Izzard, R.~G., Vink, J., \& Verbunt, F.~W. 2014,
  Astron. Astrophys., 563, A83

\bibitem[{{De Ceuster} {et~al.}(2019){De Ceuster}, Homan, Boyle, Hetherington,
  \& Yates}]{Ceuster2019a}
{De Ceuster}, F., Homan, W., Boyle, P., Hetherington, J., \& Yates, J. 2019,
  MNRAS [\eprint[arXiv]{1912.08445}]

\bibitem[{{De Marco} \& Izzard(2017)}]{DeMarco2017a}
{De Marco}, O. \& Izzard, R.~G. 2017, {Dawes Review 6: The Impact of Companions
  on Stellar Evolution}

\bibitem[{de~Val-Borro {et~al.}(2017)de~Val-Borro, Karovska, Sasselov, \&
  Stone}]{DeVal-Borro2017}
de~Val-Borro, M., Karovska, M., Sasselov, D.~D., \& Stone, J.~M. 2017, Mon.
  Not. R. Astron. Soc., 468, 3408

\bibitem[{Decin \& Al.(2020)}]{Decin2020}
Decin, L. \& Al., E. 2020, (in prep)

\bibitem[{Decin {et~al.}(2019)Decin, Homan, Danilovich, de~Koter, Engels,
  Waters, Muller, Gielen, Garc{\'{i}}a-Hern{\'{a}}ndez, Stancliffe, {Van de
  Sande}, Molenberghs, Kerschbaum, Zijlstra, \& {El Mellah}}]{Decin2019}
Decin, L., Homan, W., Danilovich, T., {et~al.} 2019, Nat. Astron.

\bibitem[{Decin {et~al.}(2006)Decin, Hony, {De Koter}, Justtanont, Tielens, \&
  Waters}]{Decin2006}
Decin, L., Hony, S., {De Koter}, A., {et~al.} 2006, A{\&}A, 456, 549

\bibitem[{Decin {et~al.}(2010)Decin, Justtanont, {De Beck}, Lombaert, {De
  Koter}, Waters, Marston, Teyssier, Sch{\"{o}}ier, Bujarrabal, Alcolea,
  Cernicharo, Dominik, Melnick, Menten, Neufeld, Olofsson, Planesas, Schmidt,
  Szczerba, {De Graauw}, Helmich, Roelfsema, Dieleman, Morris, Gallego,
  D{\'{i}}ez-Gonz{\'{a}}lez, \& Caux}]{Decin2010a}
Decin, L., Justtanont, K., {De Beck}, E., {et~al.} 2010, Astron. Astrophys.,
  521, 1

\bibitem[{Decin {et~al.}(2018)Decin, Richards, Danilovich, Homan, \&
  Nuth}]{Decin2018}
Decin, L., Richards, A.~M., Danilovich, T., Homan, W., \& Nuth, J.~A. 2018,
  Astron. Astrophys., 615

\bibitem[{Decin {et~al.}(2015{\natexlab{a}})Decin, Richards, Neufeld, Steffen,
  Melnick, \& Lombaert}]{Decin2015a}
Decin, L., Richards, A. M.~S., Neufeld, D., {et~al.} 2015{\natexlab{a}},
  Astron. Astrophys., 574, A5

\bibitem[{Decin {et~al.}(2015{\natexlab{b}})Decin, Richards, Neufeld, Steffen,
  Melnick, Lombaert, {A.M.S. Richards}, Neufeld, Steffen, G.Melnick, \&
  Lombaert}]{Decin2015}
Decin, L., Richards, A. M.~S., Neufeld, D., {et~al.} 2015{\natexlab{b}},
  Astron. Astrophys., 71-72, 87

\bibitem[{Dell'Agli {et~al.}(2015)Dell'Agli, Garc{\'{i}}a-Hern{\'{a}}ndez,
  Ventura, Schneider, {Di Criscienzo}, \& Rossi}]{DellAgli2015}
Dell'Agli, F., Garc{\'{i}}a-Hern{\'{a}}ndez, D.~A., Ventura, P., {et~al.} 2015,
  Mon. Not. R. Astron. Soc., 454, 4235

\bibitem[{D'Souza {et~al.}(2005)D'Souza, Motl, Tohline, \&
  Frank}]{DSouza:2005jx}
D'Souza, M. C.~R., Motl, P.~M., Tohline, J.~E., \& Frank, J. 2005, Astrophys.
  J., 643, 381

\bibitem[{Dullemond {et~al.}(2012)Dullemond, Juhasz, Pohl, Sereshti, Shetty,
  Peters, Commercon, \& Flock}]{Dullemond2012}
Dullemond, C.~P., Juhasz, A., Pohl, A., {et~al.} 2012, {RADMC-3D: A
  multi-purpose radiative transfer tool}

\bibitem[{Edgar(2004)}]{Edgar:2004ip}
Edgar, R.~G. 2004, New Astron. Rev., 48, 843

\bibitem[{Eggleton(1983)}]{Eggleton1983}
Eggleton, P.~P. 1983, Astrophys. J., 268, 368

\bibitem[{{El Mellah}(2016)}]{ElMellah2016}
{El Mellah}, I. 2016, PhD thesis

\bibitem[{{El Mellah} \& Casse(2015)}]{ElMellah2015}
{El Mellah}, I. \& Casse, F. 2015, Mon. Not. R. Astron. Soc., 454, 2657

\bibitem[{{El Mellah} \& Casse(2017)}]{ElMellah2016a}
{El Mellah}, I. \& Casse, F. 2017, Mon. Not. R. Astron. Soc., 467, stx225

\bibitem[{{El Mellah} {et~al.}(2019{\natexlab{a}}){El Mellah}, Sander,
  Sundqvist, \& Keppens}]{ElMellah2018}
{El Mellah}, I., Sander, A. A.~C., Sundqvist, J.~O., \& Keppens, R.
  2019{\natexlab{a}}, Astron. Astrophys., 622, A189

\bibitem[{{El Mellah} {et~al.}(2018){El Mellah}, Sundqvist, \&
  Keppens}]{ElMellah}
{El Mellah}, I., Sundqvist, J.~O., \& Keppens, R. 2018, Mon. Not. R. Astron.
  Soc., 475, 3240

\bibitem[{{El Mellah} {et~al.}(2019{\natexlab{b}}){El Mellah}, Sundqvist, \&
  Keppens}]{ElMellah2018a}
{El Mellah}, I., Sundqvist, J.~O., \& Keppens, R. 2019{\natexlab{b}}, Astron.
  Astrophys., 622, L3

\bibitem[{Falanga {et~al.}(2015)Falanga, Bozzo, Lutovinov, Bonnet-Bidaud,
  Fetisova, \& Puls}]{Falanga2015}
Falanga, M., Bozzo, E., Lutovinov, A., {et~al.} 2015, Astron. Astrophys., 577,
  A130

\bibitem[{Foglizzo {et~al.}(2005)Foglizzo, Galletti, \& Ruffert}]{Foglizzo2005}
Foglizzo, T., Galletti, P., \& Ruffert, M. 2005, Astron. Astrophys., 2201, 15

\bibitem[{Foglizzo {et~al.}(2006)Foglizzo, Galletti, Scheck, \&
  Janka}]{Foglizzo2006a}
Foglizzo, T., Galletti, P., Scheck, L., \& Janka, H.~T. 2006, Astrophys. J.,
  654, 29

\bibitem[{Frank {et~al.}(1986)Frank, King, \& Raine}]{Frank2002}
Frank, J., King, A., \& Raine, D.~J. 1986, Phys. Today, 39, 124

\bibitem[{Freytag {et~al.}(2017)Freytag, Liljegren, \&
  H{\"{o}}fner}]{Freytag2017}
Freytag, B., Liljegren, S., \& H{\"{o}}fner, S. 2017, Astron. Astrophys., 600,
  A137

\bibitem[{F{\"{u}}rst {et~al.}(2018)F{\"{u}}rst, Walton, Heida, Harrison,
  Barret, Brightman, Fabian, Middleton, Pinto, Rana, Tramper, Webb, \&
  Kretschmar}]{Fuerst2018}
F{\"{u}}rst, F., Walton, D.~J., Heida, M., {et~al.} 2018, Astron. Astrophys.,
  616, A186

\bibitem[{Gobrecht {et~al.}(2016)Gobrecht, Cherchneff, Sarangi, Plane, \&
  Bromley}]{Gobrecht2016}
Gobrecht, D., Cherchneff, I., Sarangi, A., Plane, J.~M., \& Bromley, S.~T.
  2016, Astron. Astrophys., 585, A6

\bibitem[{Groenewegen(2012)}]{Groenewegen2012}
Groenewegen, M.~A. 2012, Astron. Astrophys., 543

\bibitem[{Groenewegen {et~al.}(2002)Groenewegen, Sevenster, Spoon, \&
  P{\'{e}}rez}]{Groenewegen2002}
Groenewegen, M.~A., Sevenster, M., Spoon, H.~W., \& P{\'{e}}rez, I. 2002,
  Astron. Astrophys., 390, 511

\bibitem[{Habing \& Olofsson(2003)}]{Habing2003}
Habing, H.~J. \& Olofsson, H. 2003, Asymptot. giant branch Stars

\bibitem[{Herwig(2005)}]{Herwig2005}
Herwig, F. 2005, Annu. Rev. Astron. Astrophys., 43, 435

\bibitem[{Herwig \& Austin(2004)}]{Herwig2004}
Herwig, F. \& Austin, S.~M. 2004, Astrophys. J., 613, L73

\bibitem[{H{\"{o}}fner {et~al.}(2016)H{\"{o}}fner, Bladh, Aringer, \&
  Ahuja}]{Hofner2016}
H{\"{o}}fner, S., Bladh, S., Aringer, B., \& Ahuja, R. 2016, Astron.
  Astrophys., 594

\bibitem[{H{\"{o}}fner \& Olofsson(2018)}]{Hofner2018}
H{\"{o}}fner, S. \& Olofsson, H. 2018, {Mass loss of stars on the asymptotic
  giant branch: Mechanisms, models and measurements}

\bibitem[{Homan {et~al.}(2016)Homan, Boulangier, Decin, \& {De
  Koter}}]{Homan2016}
Homan, W., Boulangier, J., Decin, L., \& {De Koter}, A. 2016, Astron.
  Astrophys., 596 [\eprint[arXiv]{1803.05230}]

\bibitem[{Homan {et~al.}(2015)Homan, Decin, de~Koter, van Marle, Lombaert, \&
  Vlemmings}]{Homan2015}
Homan, W., Decin, L., de~Koter, A., {et~al.} 2015, Astron. Astrophys., 579,
  A118

\bibitem[{Homan {et~al.}(2018)Homan, Richards, Decin, {De Koter}, \&
  Kervella}]{Homan2018}
Homan, W., Richards, A., Decin, L., {De Koter}, A., \& Kervella, P. 2018,
  Astron. Astrophys., 616 [\eprint[arXiv]{1804.05684}]

\bibitem[{Huarte-Espinosa {et~al.}(2013)Huarte-Espinosa, Carroll-Nellenback,
  Nordhaus, Frank, \& Blackman}]{HuarteEspinosa:2012wq}
Huarte-Espinosa, M., Carroll-Nellenback, J., Nordhaus, J., Frank, A., \&
  Blackman, E.~G. 2013, Mon. Not. R. Astron. Soc., 433, 295

\bibitem[{{Iben, I.} \& Tutukov(1984)}]{IbenI.1984}
{Iben, I.}, J. \& Tutukov, A.~V. 1984, Astrophys. J. Suppl. Ser., 54, 335

\bibitem[{Jofr{\'{e}} {et~al.}(2016)Jofr{\'{e}}, Jorissen, {Van Eck}, Izzard,
  Masseron, Hawkins, Gilmore, Paladini, Escorza, Blanco-Cuaresma, \&
  Manick}]{Jofre2016}
Jofr{\'{e}}, P., Jorissen, A., {Van Eck}, S., {et~al.} 2016, Astron.
  Astrophys., 595

\bibitem[{Kamath {et~al.}(2016)Kamath, Wood, {Van Winckel}, \&
  Nie}]{Kamath2016}
Kamath, D., Wood, P.~R., {Van Winckel}, H., \& Nie, J.~D. 2016, Astron.
  Astrophys., 586, L5

\bibitem[{Khouri {et~al.}(2014)Khouri, {De Koter}, Decin, Waters, Lombaert,
  Royer, Swinyard, Barlow, Alcolea, Blommaert, Bujarrabal, Cernicharo,
  Groenewegen, Justtanont, Kerschbaum, Maercker, Marston, Matsuura, Melnick,
  Menten, Olofsson, Planesas, Polehampton, Posch, Schmidt, Szczerba,
  Vandenbussche, \& Yates}]{Khouri2014}
Khouri, T., {De Koter}, A., Decin, L., {et~al.} 2014, Astron. Astrophys., 561,
  1

\bibitem[{Khouri {et~al.}(2016)Khouri, Maercker, Waters, Vlemmings, Kervella,
  {De Koter}, Ginski, {De Beck}, Decin, Min, Dominik, O'Gorman, Schmid,
  Lombaert, \& Lagadec}]{Khouri2016}
Khouri, T., Maercker, M., Waters, L.~B., {et~al.} 2016, Astron. Astrophys.,
  591, 1

\bibitem[{Kim {et~al.}(2013)Kim, Hsieh, Liu, \& Taam}]{Kim2013}
Kim, H., Hsieh, I.~T., Liu, S.~Y., \& Taam, R.~E. 2013, Astrophys. J., 776
  [\eprint[arXiv]{1308.4140}]

\bibitem[{Kim {et~al.}(2015)Kim, Liu, Hirano, Zhao-Geisler, Trejo, Yen, Taam,
  Kemper, Kim, Byun, \& Liu}]{Kim2015a}
Kim, H., Liu, S.~Y., Hirano, N., {et~al.} 2015, Astrophys. J., 814

\bibitem[{Kim {et~al.}(2019)Kim, Liu, \& Taam}]{Kim2019}
Kim, H., Liu, S.-Y., \& Taam, R.~E. 2019, Astrophys. J. Suppl. Ser., 243, 35

\bibitem[{Kim \& Taam(2012{\natexlab{a}})}]{Kim2012a}
Kim, H. \& Taam, R.~E. 2012{\natexlab{a}}, Astrophys. J., 744
  [\eprint[arXiv]{1110.1288}]

\bibitem[{Kim \& Taam(2012{\natexlab{b}})}]{Kim2012}
Kim, H. \& Taam, R.~E. 2012{\natexlab{b}}, Astrophys. J., 759
  [\eprint[arXiv]{1209.2128}]

\bibitem[{Kim {et~al.}(2017)Kim, Trejo, Liu, Sahai, Taam, Morris, Hirano, \&
  Hsieh}]{Kim2017}
Kim, H., Trejo, A., Liu, S.~Y., {et~al.} 2017, Nat. Astron., 1, 0060

\bibitem[{Knapp {et~al.}(1998)Knapp, Young, Lee, \& Jorissen}]{Knapp1998}
Knapp, G.~R., Young, K., Lee, E., \& Jorissen, A. 1998, Astrophys. J. Suppl.
  Ser., 117, 209

\bibitem[{Lamers \& Cassinelli(1999)}]{Lamers1999}
Lamers, H. J. G. L.~M. \& Cassinelli, J.~P. 1999, Cambridge Univ. Press

\bibitem[{Levesque(2010)}]{Levesque2010}
Levesque, E.~M. 2010, New Astron. Rev., 54, 1

\bibitem[{Liljegren {et~al.}(2016)Liljegren, H{\"{o}}fner, Nowotny, \&
  Eriksson}]{Liljegren2016}
Liljegren, S., H{\"{o}}fner, S., Nowotny, W., \& Eriksson, K. 2016, Astron.
  Astrophys., 589, A130

\bibitem[{Liu {et~al.}(2017)Liu, Stancliffe, Abate, \& Matrozis}]{Liu2017}
Liu, Z.-W., Stancliffe, R.~J., Abate, C., \& Matrozis, E. 2017, Astrophys. J.,
  846, 117

\bibitem[{Lucy \& Solomon(1970)}]{Lucy1970}
Lucy, L.~B. \& Solomon, P.~M. 1970, Astrophys. J., 159, 879

\bibitem[{Maercker {et~al.}(2012)Maercker, Mohamed, Vlemmings, Ramstedt,
  Groenewegen, Humphreys, Kerschbaum, Lindqvist, Olofsson, Paladini,
  Wittkowski, de~Gregorio-Monsalvo, \& Nyman}]{Maercker2012}
Maercker, M., Mohamed, S., Vlemmings, W. H.~T., {et~al.} 2012, Nature, 490, 232

\bibitem[{Marshall {et~al.}(2004)Marshall, {Van Loon}, Matsuura, Wood,
  Zijlstra, \& Whitelock}]{Marshall2004}
Marshall, J.~R., {Van Loon}, J.~T., Matsuura, M., {et~al.} 2004, Mon. Not. R.
  Astron. Soc., 355, 1348

\bibitem[{Mastrodemos \& Morris(1999)}]{Mastrodemos1999}
Mastrodemos, N. \& Morris, M. 1999, Astrophys. J., 523, 357

\bibitem[{Moe \& {Di Stefano}(2017)}]{Moe2017}
Moe, M. \& {Di Stefano}, R. 2017, Astrophys. J. Suppl. Ser., 230, 15

\bibitem[{Mohamed \& Podsiadlowski(2007)}]{Mohamed2007}
Mohamed, S. \& Podsiadlowski, P. 2007, {Wind Roche-Lobe Overflow: a New
  Mass-Transfer Mode for Wide Binaries}

\bibitem[{Mohamed \& Podsiadlowski(2011)}]{Mohamed2011}
Mohamed, S. \& Podsiadlowski, P. 2011, ASP Conf. Ser., 445

\bibitem[{Nicholls {et~al.}(2010)Nicholls, Wood, \& Cioni}]{Nicholls2010}
Nicholls, C.~P., Wood, P.~R., \& Cioni, M.-R.~R. 2010, Mon. Not. R. Astron.
  Soc., 405, 1770

\bibitem[{Nordhaus \& Blackman(2006)}]{Nordhaus2006}
Nordhaus, J. \& Blackman, E.~G. 2006, {Low-mass binary-induced outflows from
  asymptotic giant branch stars}

\bibitem[{Paczynski(1977)}]{Paczynski1977}
Paczynski, B. 1977, Astrophys. J., 216, 822

\bibitem[{Pavlovskii {et~al.}(2017)Pavlovskii, Ivanova, Belczynski, \&
  Van}]{Pavlovskii2017}
Pavlovskii, K., Ivanova, N., Belczynski, K., \& Van, K.~X. 2017, Mon. Not. R.
  Astron. Soc., 465, 2092

\bibitem[{Puls {et~al.}(2008)Puls, Vink, \& Najarro}]{Puls2008}
Puls, J., Vink, J.~S., \& Najarro, F. 2008, Astron. Astrophys. Rev., 16, 209

\bibitem[{Quaintrell {et~al.}(2003)Quaintrell, Norton, Ash, Roche, Willems,
  Bedding, Baldry, \& Fender}]{Quaintrell2003}
Quaintrell, H., Norton, A.~J., Ash, T. D.~C., {et~al.} 2003, Astron.
  Astrophys., 401, 313

\bibitem[{Quast {et~al.}(2019)Quast, Langer, \& Tauris}]{Quast2019a}
Quast, M., Langer, N., \& Tauris, T.~M. 2019, Astron. Astrophys., 628, A19

\bibitem[{Ramos-Larios {et~al.}(2016)Ramos-Larios, Santamar{\'{i}}a, Guerrero,
  Marquez-Lugo, Sabin, \& Toal{\'{a}}}]{Ramos-Larios2016}
Ramos-Larios, G., Santamar{\'{i}}a, E., Guerrero, M.~A., {et~al.} 2016, Mon.
  Not. R. Astron. Soc., 462, 610

\bibitem[{Ramstedt {et~al.}(2014)Ramstedt, Mohamed, Vlemmings, Maercker,
  Montez, Baudry, {De Beck}, Lindqvist, Olofsson, Humphreys, Jorissen,
  Kerschbaum, Mayer, Wittkowski, Cox, Lagadec, Leal-Ferreira, Paladini,
  P{\'{e}}rez-S{\'{a}}nchez, \& Sacuto}]{Ramstedt2014}
Ramstedt, S., Mohamed, S., Vlemmings, W.~H., {et~al.} 2014, Astron. Astrophys.,
  570, 10

\bibitem[{Ramstedt {et~al.}(2009)Ramstedt, Sch{\"{o}}ier, \&
  Olofsson}]{Ramstedt2009}
Ramstedt, S., Sch{\"{o}}ier, F.~L., \& Olofsson, H. 2009, Astron. Astrophys.,
  499, 515

\bibitem[{Ricker \& Taam(2012)}]{Ricker2012}
Ricker, P.~M. \& Taam, R.~E. 2012, Astrophys. J., 746, 74

\bibitem[{Ruffert(1999)}]{Ruffert1999}
Ruffert, M. 1999, Astron. Astrophys., 06, 17

\bibitem[{Saladino {et~al.}(2019)Saladino, Pols, \& Abate}]{Saladino2019}
Saladino, M.~I., Pols, O., \& Abate, C. 2019, eprint arXiv:1903.04515
  [\eprint[arXiv]{1903.04515}]

\bibitem[{Sander {et~al.}(2017)Sander, F{\"{u}}rst, Kretschmar, Oskinova, Todt,
  Hainich, Shenar, \& Hamann}]{Sander2017}
Sander, A. A.~C., F{\"{u}}rst, F., Kretschmar, P., {et~al.} 2017, Astron.
  Astrophys. Vol. 610, id.A60, 19 pp., 610, A60

\bibitem[{Sargent {et~al.}(2011)Sargent, Srinivasan, \& Meixner}]{Sargent2011}
Sargent, B.~A., Srinivasan, S., \& Meixner, M. 2011, Astrophys. J., 728, 93

\bibitem[{Shklovsky(1956)}]{Shklovsky1956}
Shklovsky, I.~S. 1956, {The nature of planetary nebulae and their nuclei}

\bibitem[{Straniero {et~al.}(1997)Straniero, Chieffi, Limongi, Busso, Gallino,
  \& Arlandini}]{Straniero1997a}
Straniero, O., Chieffi, A., Limongi, M., {et~al.} 1997, Astrophys. J., 478, 332

\bibitem[{Tauris \& van~den Heuvel(2003)}]{Tauris2003}
Tauris, T.~M. \& van~den Heuvel, E. 2003, Cambridge Astrophys. Ser.
  [\eprint[arXiv]{0303456}]

\bibitem[{Teunissen \& Keppens(2019)}]{Teunissen2019}
Teunissen, J. \& Keppens, R. 2019, Comput. Phys. Commun., 245, 106866

\bibitem[{Toro {et~al.}(1994)Toro, Spruce, \& Speares}]{Toro1994}
Toro, E.~F., Spruce, M., \& Speares, W. 1994, Shock Waves, 4, 25

\bibitem[{Vreugdenhil \& Koren(1993)}]{Vreugdenhil1993}
Vreugdenhil, C.~B. \& Koren, B. 1993, {Numerical methods for
  advection--diffusion problems} (Vieweg), 373

\bibitem[{Webbink(1984)}]{Webbink1984}
Webbink, R.~F. 1984, Astrophys. J., 277, 355

\bibitem[{Xia {et~al.}(2018)Xia, Teunissen, {El Mellah}, Chan{\'{e}}, \&
  Keppens}]{Xia2017}
Xia, C., Teunissen, J., {El Mellah}, I., Chan{\'{e}}, E., \& Keppens, R. 2018,
  Astrophys. J. Suppl. Ser., 234, 30

\bibitem[{Yang {et~al.}(2018)Yang, Bonanos, Jiang, Gao, Xue, Wang, Lam,
  Spetsieri, Ren, \& Gavras}]{Yang2018}
Yang, M., Bonanos, A.~Z., Jiang, B.~W., {et~al.} 2018, Astron. Astrophys., 616,
  A175

\bibitem[{Zahn(1977)}]{Zahn1977}
Zahn, J.-P. 1977, Astron. Astrophys., 57, 383

\bibitem[{Zanni \& Ferreira(2013)}]{Zanni2013}
Zanni, C. \& Ferreira, J. 2013, Astron. Astrophys., 550, A99

\end{thebibliography}
\end{tiny}

\begin{appendix}

\section{Appendix A: orbital separation changes due to mass-loss}
\label{app:app3}

\begin{figure}
\begin{subfigure}{\columnwidth}
  \centering
% 1col
\includegraphics[width=0.99\columnwidth]{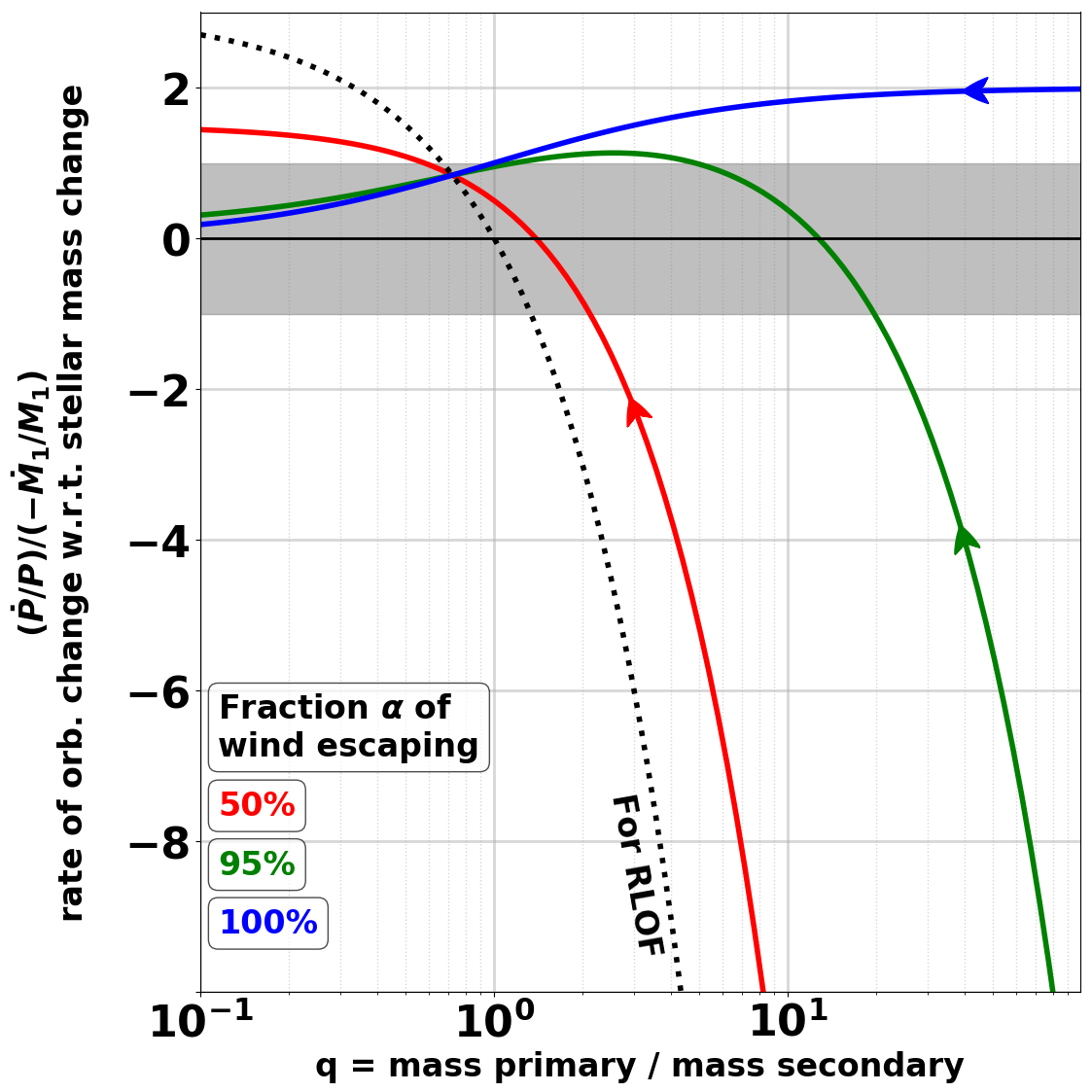}
\end{subfigure}
\begin{subfigure}{\columnwidth}
  \centering
% 1col
\includegraphics[width=0.99\columnwidth]{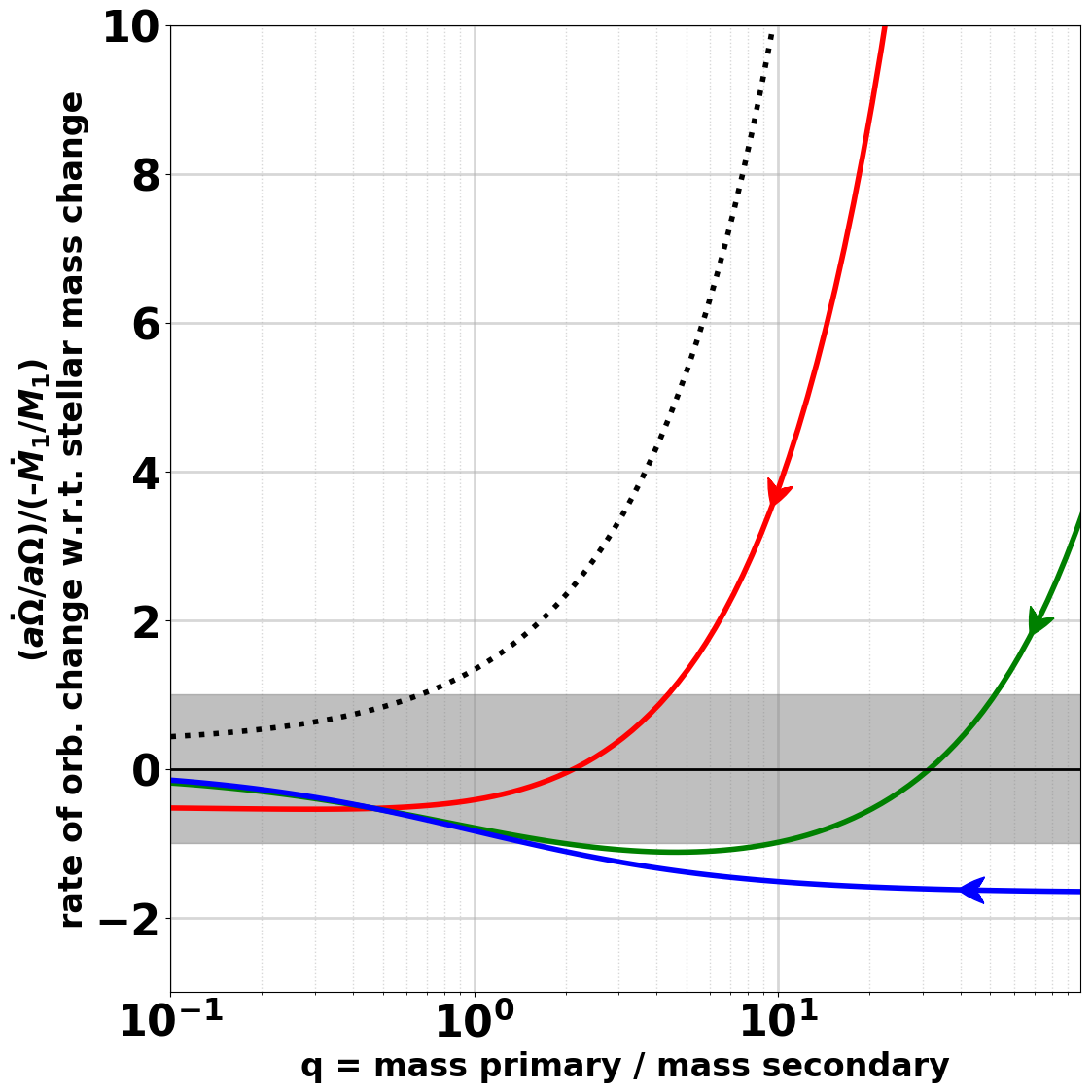}
\end{subfigure}
\caption{Rate of change of the orbital period (upper panel) and orbital speed $a\Omega$ (bottom panel) compared to the rate of stellar mass change. In the grey shaded region, the mass of the donor star changes faster than the orbital period or speed. The two limit cases are conservative mass transfer (dotted line, RLOF) and pure mass-loss without accretion by the secondary (solid blue line). In-between, the green and red solid lines are for an accretion efficiency by the secondary of 5\% and 50\% respectively. The arrows indicate that as mass transfer proceeds, the mass ratio can only decrease.}
\label{fig:orb_sep_decay}
\end{figure}

If the orbital separation increases as the system looses mass, the orbital period increases, and if the orbital period decreases as the system looses mass, the orbital separation decreases. However, mixed cases can occur. Since the total mass of the system diminishes, if the orbital separation decreases slowly, the orbital period can increase. Comparing Figure\,\ref{fig:orb_decay} and the upper panel in Figure\,\ref{fig:orb_sep_decay} confirms that when the mass ratio is such that $(\dot{a}/a)/(-\dot{M}_1/M_1)$ is negative (\ie the orbit is shrinking) but close to zero, the orbital period can still be increasing. The net effect on the orbital speed is visible in the bottom panel in Figure\,\ref{fig:orb_sep_decay}.

\end{appendix}

\end{document}